\documentclass[aps,prd,reprint,longbibliography,nofootinbib]{revtex4-2}

\usepackage{lmodern}
\usepackage[T1]{fontenc}
\usepackage{amsmath}
\usepackage{amssymb}

\usepackage{url}
\usepackage{multirow}
\usepackage{caption}
\usepackage[labelformat=simple]{subcaption}

\usepackage[nodayofweek]{datetime}
\usepackage{enumerate}
\usepackage{xspace}
\usepackage{xcolor}
\usepackage{graphicx}
\usepackage[pdftex,hidelinks]{hyperref}
\usepackage{bookmark}
\usepackage[compat=1.1.0]{tikz-feynman}
\usepackage{placeins}
\usepackage{float}
\usepackage[capitalise]{cleveref}

\usepackage{import}
\usepackage{tabularx, booktabs}
\newcolumntype{Y}{>{\centering\arraybackslash}X}

\usepackage{siunitx}

\graphicspath {{figures/}}

\newcommand{\diff}{\ensuremath{\mathop{}\!\mathrm{d}}\xspace}
\newcommand{\x}{\ensuremath{\boldsymbol{x}}\xspace}

\newcommand{\cond}{\ensuremath{\boldsymbol{y}}\xspace}  

\newcommand{\pt}{\ensuremath{p_\mathrm{T}}\xspace}

\newcommand{\pcjedi}{\mbox{PC-JeDi}\xspace}
\newcommand{\jeditwo}{\mbox{PC-Droid}\xspace}

\newcommand{\nvidia}{\mbox{NVIDIA\textsuperscript{\textregistered}}\xspace}

\begin{document}
  \title{\jeditwo: Faster diffusion and improved quality for particle cloud generation}
  \author{Matthew Leigh}
  \email{matthew.leigh@unige.ch}
  \author{Debajyoti Sengupta}
  \email{debajyoti.sengupta@unige.ch}
  \author{John Andrew Raine}
  \email{john.raine@unige.ch}
  \author{Guillaume Quétant}
  \author{Tobias Golling}
  \affiliation{Département de physique nucléaire et corpusculaire, University of Geneva, Switzerland}

  \begin{abstract}
Building on the success of \pcjedi we introduce \jeditwo, a substantially improved diffusion model for the generation of jet particle clouds.
By leveraging a new diffusion formulation, studying more recent integration solvers, and training on all jet types simultaneously, we are able to achieve state-of-the-art performance for all types of jets across all evaluation metrics.
We study the trade-off between generation speed and quality by comparing two attention based architectures, as well as the potential of consistency distillation to reduce the number of diffusion steps.
Both the faster architecture and consistency models demonstrate performance surpassing many competing models, with generation time up to two orders of magnitude faster than \pcjedi and three orders of magnitude faster than Delphes.
  \end{abstract}

  \maketitle

  \section{Introduction}
  An incredible amount of computing resources is required to keep up with demands for simulated events at the intensity frontier of high energy physics~(HEP).
As such, focus has turned to fast surrogate models for event and detector simulation.
Deep generative models and modern machine learning techniques have shown great promise in improving the fidelity and speed of fast simulation approaches~\cite{deOliveira:2017rwa,Paganini:2017dwg,Paganini:2017hrr,Erdmann:2018jxd,Belayneh:2019vyx,Buhmann:2020pmy,Krause:2021ilc,Krause:2021wez,atlfast3,ATLAS:2022jhk,Adelmann:2022ozp,Krause:2022jna,Liu:2023lnn,Otten:2019hhl,Hashemi:2019fkn,DiSipio:2019imz,Butter:2019cae,ArjonaMartinez:2019ahl,Gao:2020zvv,Alanazi:2020klf,Bellagente:2020piv,Velasco:2020nqr,Butter:2020tvl,Howard:2021pos,Quetant:2021hgi}.
Recently, diffusion models have come to the fore in a wide range of disciplines, notably for images generation~\cite{ClassifierFreeGuidance, Dalle2, DDPM, ImprovedDDPM, DDIM, Karras2022}.
However, they have also demonstrated promise in applications in HEP for fast detector simulation~\cite{CaloScore,CaloClouds,Acosta:2023zik}, the generation of jets~\cite{PCJedi,FPCD,Butter:2023fov}, unfolding~\cite{Shmakov:2023kjj}, and anomaly detection~\cite{Mikuni:2023tok}. 

The recent rise of diffusion models has also lead to a rapid development of approaches~\cite{yang2023diffusion}, similar to the development observed in generative adversarial networks~\cite{gui2020review}.
As such, the level of performance is continuously improving with more stable training procedures and improved differential equation solvers for generation.

In this work we introduce \jeditwo (Particle Cloud Diffusion of reconstructed objects with improved denoising), a significant improvement over the \pcjedi approach introduced in Ref.~\cite{PCJedi}, demonstrating both a decreased inference time and higher-fidelity generation.
We use a more recently developed formulation for diffusion models~\cite{Karras2022} and combine it with improved diffusion sampling algorithms \cite{DPMSolver,DPMSolverPP}.
We compare and contrast different network architectures, balancing the overall performance against generation time.
Furthermore, in an effort to greatly reduce the generation time we study the application of consistency models~\cite{song2023consistency} for jet generation.

With \jeditwo we achieve state-of-the-art performance as measured on standard metrics, as well as a new set of benchmarks.
The repository\footnote{\url{https://github.com/rodem-hep/pcdroid}} used in this work is publicly available.


  \section{Method}
  \jeditwo is a family of models which are trained to generate the constituent particles of jets either conditionally, given the desired kinematics of the jet, or unconditionally, with jet kinematics sampled from distributions learned from the training dataset.

\subsection{Improved diffusion formulation}

One of the key developments in \jeditwo is the change in diffusion paradigm and setup.
\jeditwo follows the `EDM' noise scheduler and network preconditioning from Ref.~\cite{Karras2022}.
This introduces several crucial changes.

First, following the notation from Ref.~\cite{PCJedi}, the signal and noise rates are set to $\gamma(t)=1$ and $\sigma(t)=t$.
This simplifies the stochastic differential equation (SDE) to
\begin{equation*}
    \diff x_t = \sqrt{2t}\diff w,
\end{equation*}
with the corresponding reverse ordinary differential equation (ODE) reducing to
\begin{equation*}
    \diff x_t = - t \nabla_x \log p(x; t) \diff t.
\end{equation*}
Consequently, the ODE has solutions with straighter trajectories.
This results in fewer truncation errors during generation compared to the previous variance preserving approach.
The significant benefit of this change is being able to perform the reverse diffusion process in fewer steps.
In addition, $t$ and $\sigma$ are now interchangeable, with $t$ no longer bounded between 0 and 1 and instead in the range $t = \sigma \in [0, 80)$.

Second, during training we sample the noise rates using a log-norm distribution defined by ${\log(\sigma)\sim\mathcal{N}\left(-1.2, 1.2\right)}$.
At low values of $\sigma$, the magnitude of the artificial noise added to the samples approaches the natural stochasticity in the data.
At high values of $\sigma$, the denoising targets produce an estimator of the score function with excessively high variance.
The log-norm sampling adds focus to intermediate values of $\sigma$, where the most gain in performance is possible.
The sampling distribution is taken from Ref.~\cite{Karras2022}, and is found to be similarly optimal for our application.

Third, smarter skip and scaling connections are introduced to the network architecture.
The variables $c_\mathrm{in}(\sigma)$, $c_\mathrm{out}(\sigma)$ and $c_\mathrm{skip}(\sigma)$ are all parametrised by the noise strength $\sigma$, and are combined for the denoised estimate
\begin{equation*}
    D_\theta(\x;\sigma, \cond) = c_\mathrm{out} F_\theta(c_\mathrm{in}\x;\sigma, \cond)+c_\mathrm{skip}\x,
\end{equation*}
where $F_\theta$ is the prediction from the network for the input point cloud \x and conditional parameters \cond.
As the variance is no longer preserved during the diffusion process, the scaling functions are used to maintain the unit variance of the raw inputs and outputs of the network.
The skip connection allows the $\x$ to bypass the network at low $\sigma$, as the input is closer to the target.

Finally, this new framework results in a change to the objective function of the network.
Here, the loss is calculated from the distance between the denoised output and the true target $\x_0$
\begin{equation*}
    \mathcal{L} = ||D_\theta(\x;\sigma, \cond)  - \x_0||^2_2,
\end{equation*}
where previously, the loss was calculated from the difference in the predicted noise added in \x.

\begin{figure}[htpb]
    \centering
    \includegraphics[width=0.5\textwidth]{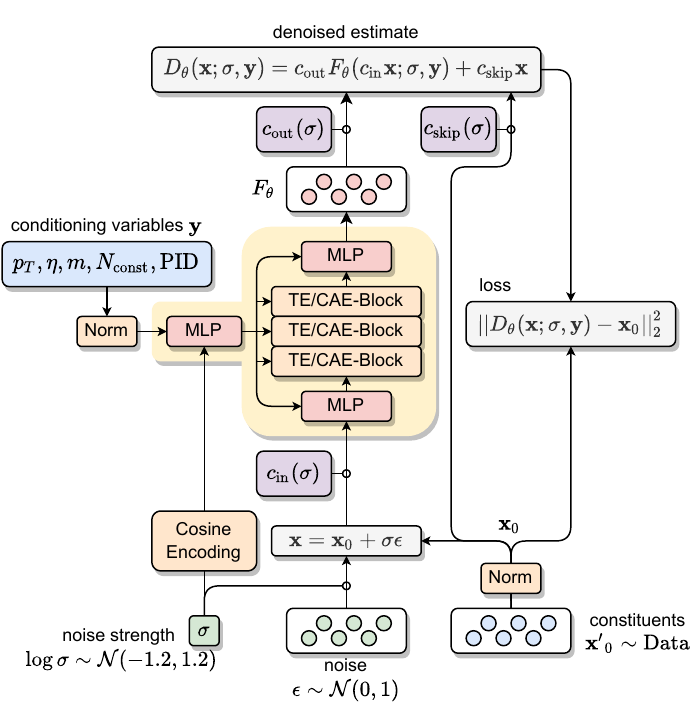}
    \caption{\jeditwo network architecture and training setup.}
    \label{fig:general_arch_train}
\end{figure}

The general \jeditwo architecture and the training process is shown in \cref{fig:general_arch_train}.

Another key development has been in the choice of integration solvers used during inference.
Several state-of-the-art algorithms have been studied including those provided by \texttt{k-diffusion} library.%
\footnote{\url{https://github.com/crowsonkb/k-diffusion/tree/v0.0.15}}
The most promising solvers are presented in this work.

In \pcjedi dedicated models were trained for each jet type; in \jeditwo a single conditional model for all five particle types (PID) ($q$ - light quark; $g$ - gluon; $t$ - top quark; $W$ - $W$ boson; $Z$ - $Z$ boson) is produced.
By training simultaneously on all jet types for a single conditional model, we are able to increase the overall training statistics for the \jeditwo model.
This is observed to result in improved generation quality for all jet types in comparison to individual models trained separately for each.

\subsection{Model training}

A denoising transformer is trained for each of the JetNet datasets introduced in Ref.~\cite{MPGAN} comprising of up to 30~\cite{JetNet} and 150~\cite{JetNet150} constituents.
To allow comparisons to \pcjedi, the optimiser, learning rate scheduler, and most of the hyperparameters are left unchanged.
The only notable changes are the reduction in the number of transformer encoder layers from four to three, and the inclusion of an extra MLP to combine the cosine embedding of $\sigma$ with the context vector.
For the 150 constituents model we double the token dimension and the width of the MLP layers.
As in \pcjedi, each jet constituent is represented by its coordinates relative to the jet centre $\Delta \eta$ and $\Delta \phi$, and its transverse momentum of the form $\log(\pt + 1)$.

  \subsection{Cross-attention encoder}




In \pcjedi only a self-attention transformer encoder (TE) architecture was studied which, although very expressive, is computationally expensive.
The number of operations scales with $\mathcal{O}(N_{const}^2)$ for the number of constituents $N_{const}$.
As diffusion models require many network passes during generation, this makes self-attention a suboptimal choice for fast generation.
For the 30 constituent models this does not present a large problem, however when moving to 150 constituents the impact is non-negligible.
Therefore, in addition to the transformer model we also introduce the cross-attention encoder (CAE) as a faster and more memory efficient permutation equivariant network.

\begin{figure}[htpb]
    \centering
    \includegraphics[width=0.5\textwidth]{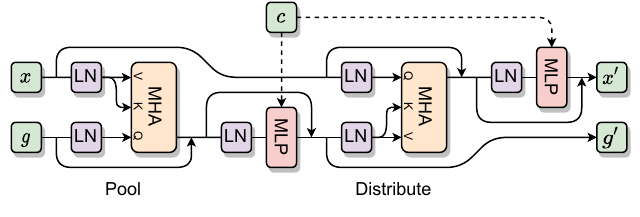}
    \caption{A single cross-attention encoder block updating both an input point cloud $x$ and global tokens $g$, using layer-norm (LN), multi-headed attention (MHA). Converging arrows represent vector addition and contextual information $c$ is injected into the network by concatenating to the inputs of the MLPs. The first attention operation effectively pools information from the point cloud into the global tokens. The second attention operation is inverted, and information from the updated globals token are distributed back to the point cloud.}
    \label{fig:cae_network}
\end{figure}

A schematic overview of the CAE-Block is shown in \cref{fig:cae_network}.
In a CAE-Block, the input point cloud is used to update a group of global tokens using multi-headed cross-attention.
The number of global tokens $N_{global}$ is a hyperparameter.
The global tokens are then further updated using a residual multi-layer perceptron (MLP), before being redistributed back to the point cloud using another cross-attention layer and a residual MLP update.
This process of global pooling followed by distribution can be thought of as a transformer analogue to EPiC layers~\cite{EPiC-GAN}.
We construct the full CAE by stacking together multiple CAE-Blocks in sequence, and allowing the initial set of global tokens to be fully learnable.
If $N_{global}=1$, the distribution attention operation utilises a sigmoid instead of a softmax function.
The number of attention operations for the CAE scales with $\mathcal{O}(N_{const} \times N_{global})$.

We train two models using CAE layers, with \mbox{$N_{global}=1$} or 16, on the 150 constituent dataset using the same training setup, model dimension, and number of layers as the baseline transformer for a fair comparison.

  \subsection{Consistency models}
  In this work we investigate the process of consistency distillation (CD) \cite{song2023consistency}, whereby a teacher network is used to train a student network in order to solve the reverse diffusion process in less time steps, even enabling \textit{one-shot} generation.
The teacher network in this case is a standard diffusion model which can already be used in conjunction with an integration method to solve the reverse diffusion ODE \cite{Song2020}.

During training, the objective of the student network is to map all points sampled along a same ODE trajectory to the same output.
To ensure that the student model does not collapse to a constant function, a boundary condition is enforced such that the it approaches the identity map as $\sigma$ approaches zero.
We follow the approach in Ref.~\cite{song2023consistency} and use skip connections to enforce this.
This boundary condition means that the global minimum of the training loss is reached only when the student network learns to map each point of the ODE trajectory to its endpoint.

At the start of training the student model is initialised with the same parameter values as the teacher model.
During training, two adjacent points on the ODE are sampled using the teacher network and sampler, and are subsequently evaluated with the student.
The training loss is the mean squared distance between these two outputs.
To stabilise training, the student network is duplicated into a online and target network, a process commonly found in deep reinforcement learning \cite{mnih2013playing, lillicrap2016target}.
Gradients are only propagated through the online network which always processes the ODE sample with the larger $\sigma$.
After each iteration, the target network is synced with the student network using an exponentially decaying average of the parameters.

Although both are distillation methods, consistency models differ from progressive distillation models~(PD)~\cite{salimans2022progressive}, such as the approach applied to jet generation in Ref.~\cite{FPCD}.
In progressive distillation, new models are iteratively trained to predict the amount of noise removed in $N$ steps of the nominal model.
This processes is repeated with each new model trained to reduce the overall number of steps required, culminating in a faster generation time.
No information about the ODE trajectories are required, only the total amount of noise removed after $N$ steps.

  \subsection{Unconditional generation}
  \jeditwo is trained as a conditional generative model, with the jet kinematics $(\pt, \eta, m)$, the number of constituents ($N_{const}$), and the PID used as conditions.
To allow for unconditional generation we present a method using normalizing flows \cite{rezende2015variational} to generate these variables as an initial step.

One normalizing flow (Flow-$\vec{p}$) is trained to learn the probability density $p(\pt, \eta, m, N_{const} |\textrm{PID})$.
The transformer diffusion models are parametrised by the number of constituents, and therefore this value is always required at generation time.
The number of constituents is correlated to many of the jet properties, and thus we require a normalizing flow to learn these correlations.
A second conditional normalizing flow (Flow-$N$) is trained to learn the probability density $p(N_{const}|\pt, \eta, m, \textrm{PID})$.
This second flow is designed to highlight how \jeditwo can be used as a surrogate fast simulator where one would typically know the parton type and kinematics, but not the number of particles present in the final jet.

Both flows are trained with a maximum log likelihood objective, and consist of five transformation layers using rational quadratic splines~\cite{NeuralSplines}.
Coupling layers are used in the flow generating values for $p(\pt, \eta, m, N_{const} |\textrm{PID})$, whilst the one dimensional flow for p($N_{const}|\pt, \eta, m, \textrm{PID})$ simply learns a single spline per layer.
For both flows, a dequantisation step is used to simplify the learning of the $N_{const}$ distribution; at inference time the generated value is rounded to the nearest integer.
We use the Adam \cite{Adam} optimiser with default settings with a cosine learning rate.
As these distributions are not very complex, near perfect performance is obtained using only four layers and 100k learnable parameters.

This approach is preferential over implicitly unconditional models as it segments what the model needs to learn along a logical line.
The kinematics of a jet is strongly correlated to and influential on its substructure, and therefore the kinematics of the individual constituents.
Providing this information to the diffusion model is observed to improve the overall performance.
Secondly, the kinematics can be far more easily learned and modelled with a normalizing flow than embedding it in the same network.

  \subsection{Comparison to prior work}
  Diffusion models are a fast growing family of generative models, and are becoming more actively studied for their use in high energy physics.
Whilst this work was being undertaken, Ref.~\cite{FPCD} introduced a fast diffusion model for jet particle cloud generation exploiting progressive distillation.

We are similar to FPCD in that we train on all jet types simultaneously, have a second model for generating the conditions, and can generate jets with up to 150 constituents.
However, our approach differs as we consider not only transformers but also introduce a faster, novel architecture in comparison, we focus on generation speed as well as performance, and study the use of consistency distillation.
We also use a different diffusion paradigm to train and perform inference.
The additional conditioning variables are also modelled differently between the two approaches.
In Ref.~\cite{FPCD} a second diffusion model is jointly trained to generate the kinematics and $N_{const}$ of the jets given the PID.
However, generating the conditional information is an orthogonal task to the constituent generation, and we find we do not need to train both jointly as there are no shared weights in the two networks.
Using a normalizing flow is also significantly faster at inference than a second diffusion model, and well suited to the structured vector output.
In addition, it is unlikely that a fast surrogate model will be used solely for the amplification of statistics from the training distribution and with conditional generation any desired distributions over the jet kinematics can be achieved.

Other adversarial approaches such as MPGAN~\cite{MPGAN} and EPiC-GAN~\cite{EPiC-GAN} rely on implicitly learning the correlations to the jet kinematics without a second network.
In EPiC-GAN a kernel-density estimator~(KDE) is used to model the number of constituents, which is a conditional parameter of the generation and correlated to the jet invariant mass and \pt.
However, the KDE does not take into account correlations between the number of constituents and the kinematics of the jet.

Conditioning is also studied in JetFlow~\cite{JetFlow}, where normalizing flows are used to generate jets.
Due to the nature of normalizing flows, in comparison to the diffusion and GAN based architectures, JetFlow is not permutation invariant.
Furthermore, as normalizing flows have a fixed dimensionality, jets with fewer than 30 particles are zero-padded up to the maximum size.
Small levels of noise added to the empty constituents during training.

Due to the fixed size of normalizing flows, jets with fewer than 30 particles are zero-padded up to the maximum size, with small levels of noise added to the empty constituents.
The number of constituents, along with the jet mass, are provided to the network as conditions during training and at inference.
To generate new jets with JetFlow, the conditioning variables are sampled sequentially from two cumulative distribution functions~(CDF), with a separate jet mass CDF for each value of $N_{const}$.

Concurrent with this work, Ref.~\cite{Kach:2023rqw} studied the use of similar cross-attention layers for particle cloud generation.
A distinguishing feature of the CAE layer is that is extends to multiple global tokens and utilises attention to redistribute them back to the point cloud, while in Ref.~\cite{Kach:2023rqw} distribution is performed by concatenation to the individual points.

  \section{Results}
  \subsection{JetNet30}

To assess the quality of the generated jets we compare distributions of several observables to the JetNet30 test set, which we label as Delphes.
For a quantitative analysis, we use the metrics\footnote{For metrics which require bootstrapping, we use 16 batches of 20000 events for all methods.} introduced in Ref.~\cite{PCJedi} and Ref.~\cite{MPGAN}.
For each metric we establish an ideal limit by comparing the training and test sets, which corresponds to the natural variation in the Delphes samples.
All \jeditwo samples used in this study are taken from the fully conditional generation regime.
We observe very little difference in performance between sampling all conditions from the normalizing flow for each jet, and taking the jet kinematics from the training data.

\begin{figure*}[htpb]
    \centering
        \includegraphics[width=0.3\linewidth]{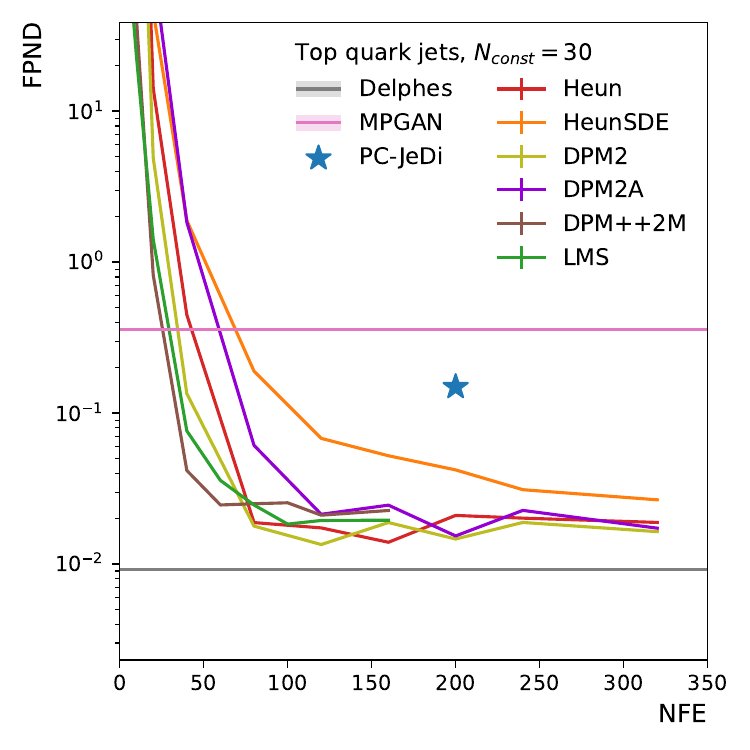}
        \includegraphics[width=0.3\linewidth]{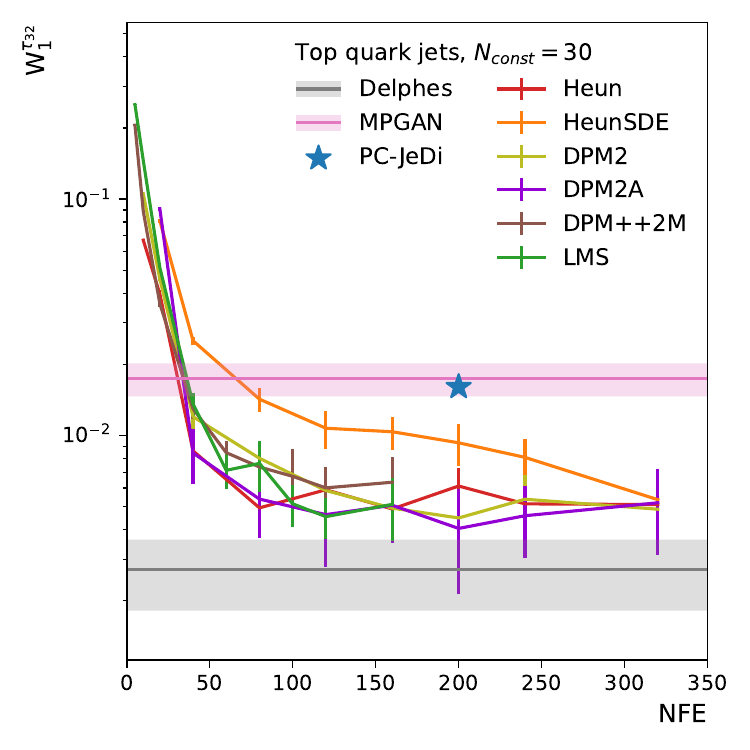}
        \includegraphics[width=0.3\linewidth]{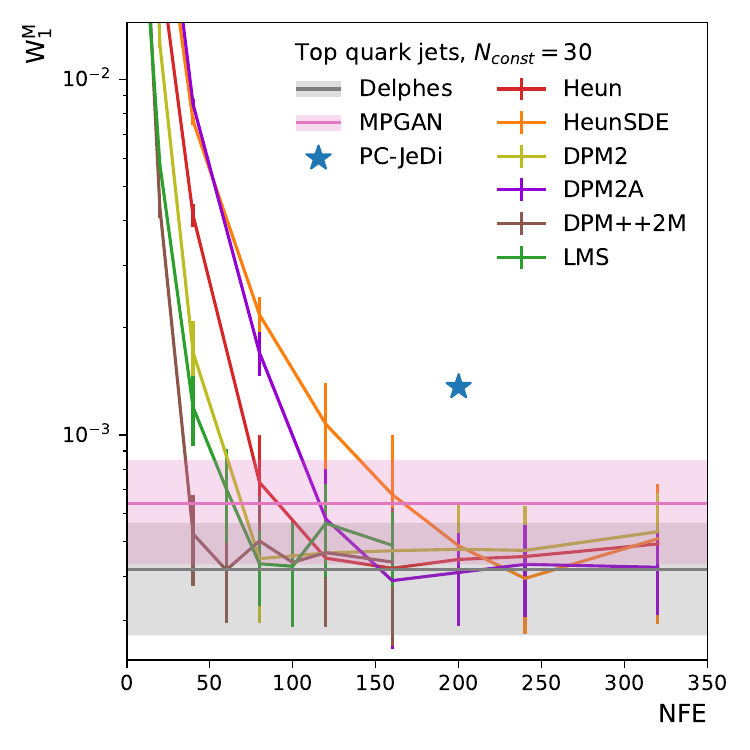}
        \caption{Performance as measured by FPND (left), W$_1^{\tau_{32}}$ (middle), and W$_1^\mathrm{M}$ (right) on the NFE for top jets with up to 30 constituents.
    We generate samples with \jeditwo using Heun and HeunSDE methods~\cite{Karras2022}, DPM2~\cite{DPMSolver}, DPM2 with ancestral sampling (DPM2A), DPM++2M~\cite{DPMSolverPP}, and LMS~\cite{ODEBook}.
    }
    \label{fig:metrics_vs_steps-30}
\end{figure*}

We test a wide variety of integration solvers in the generation stage for \jeditwo.
These include a fourth order linear multistep method~(LMS)~\cite{ODEBook} and DPM-Solver-2~(DPM2) \cite{DPMSolver}.
We study the trade-off in quality of the generated jets and the amount of neural function evaluations (NFE) in order to choose the best solver and a corresponding optimal step size.
For this comparison we look at the FPND, $\mathrm{W}_1^M$, and the $\mathrm{W}_1^{\tau_{32}}$ metrics.
Results for \pcjedi are obtained using the Euler-Maruyama sampler at 200 NFE~\cite{PCJedi}.
From \cref{fig:metrics_vs_steps-30}, we see most solvers saturate at around $100$ NFE.
While there is no clearly superior method, we observe that the LMS solver performs the best across most metrics and henceforth we use the LMS solver at 100 NFE for all \jeditwo results.
We also find that with most solvers \jeditwo outperforms MPGAN with as few as $20$ steps.

\begin{figure*}[htpb]
    \centering
        \includegraphics[width=0.9\linewidth]{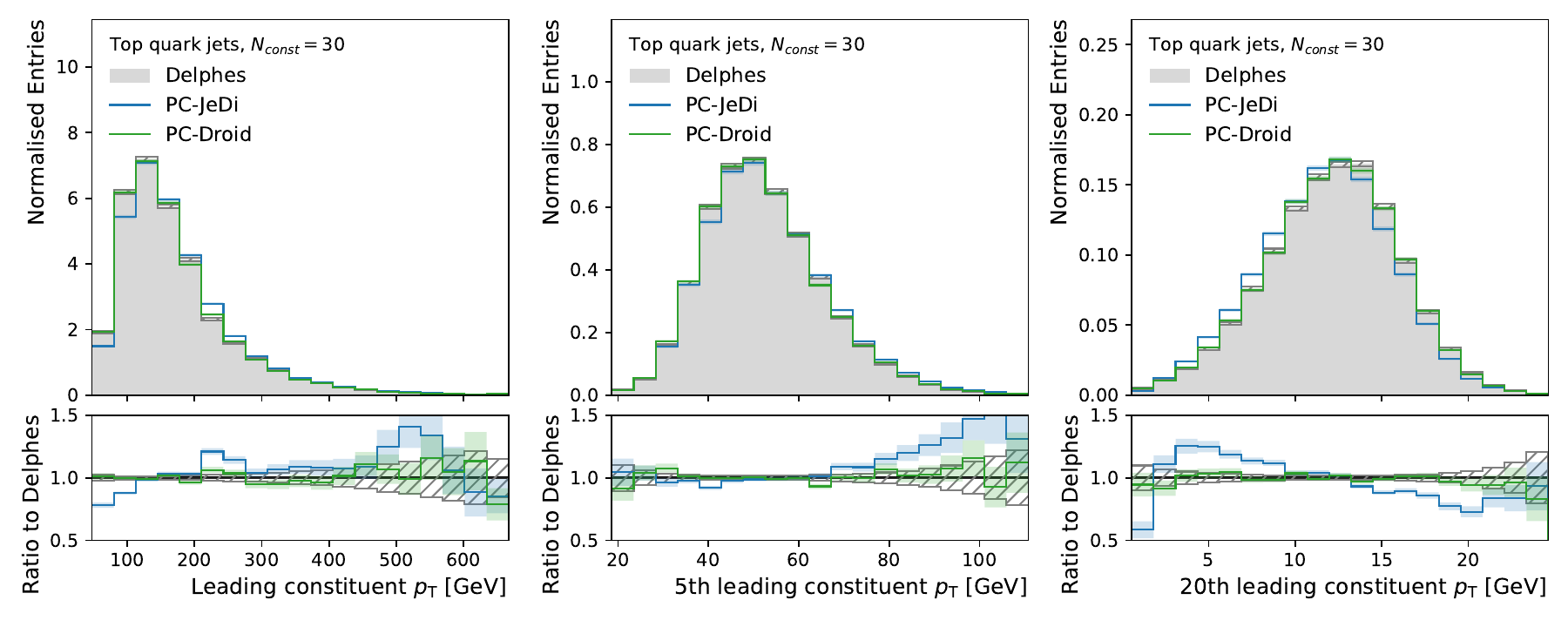} \\
        \includegraphics[width=0.9\linewidth]{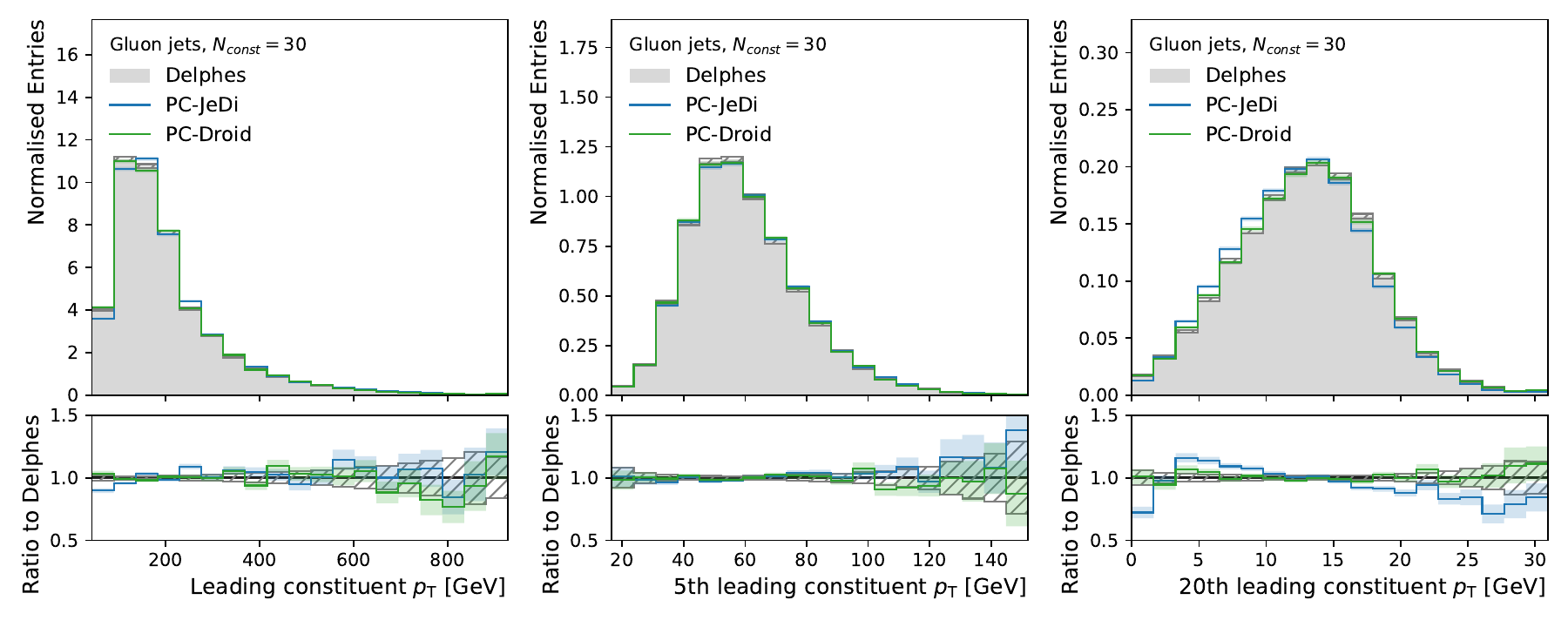}
    \caption{Comparison of \pt distributions of the leading, fifth leading, and twentieth leading constituents of the generated top and gluon jets with up to 30 constituents.}
    \label{fig:const-pt_dist-30}
\end{figure*}

Accurately modelling individual constituents is a crucial requirement when it comes to jet generation.
\Cref{fig:const-pt_dist-30} shows the \pt distributions of the leading, fifth leading, and twentieth leading constituents of the generated top and gluon jets as modelled by \jeditwo and \pcjedi.
\jeditwo demonstrates improved agreement with Delphes compared to \pcjedi across all the constituents, especially in the tails.

\begin{figure*}[htpb]
    \centering
        \includegraphics[width=0.49\linewidth]{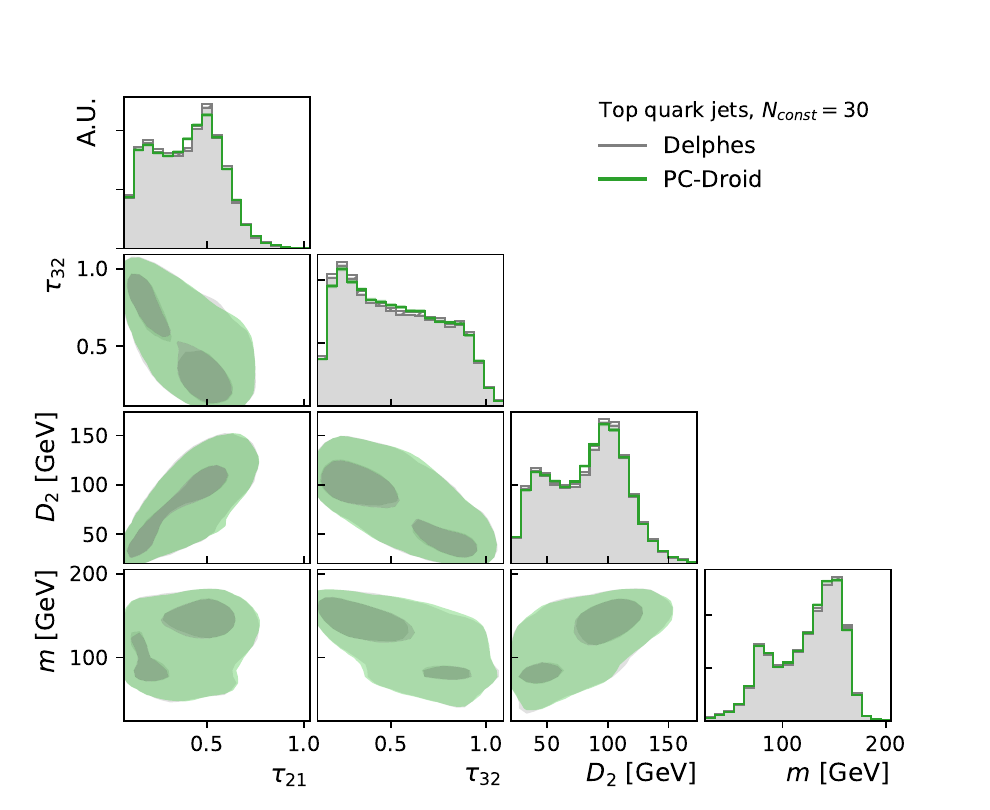}
        \includegraphics[width=0.49\linewidth]{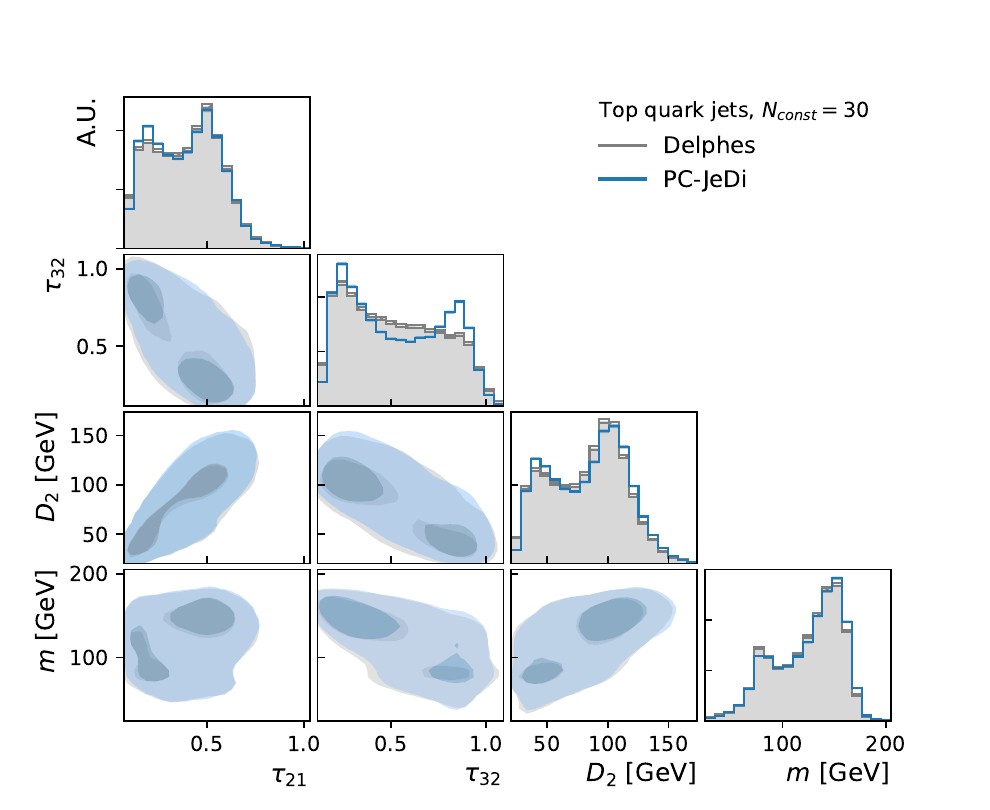}
    \caption{
        Mass and substructure distributions of the generated top jets with up to 30 constituents.
        The diagonal consists of the marginals of the distributions and the off-diagonal elements contain the joint distributions.
        }
    \label{fig:hlvs-30}
\end{figure*}

Next, we look at the substructure variable distributions of the generated jets and correlations between them, which are crucial for jet tagging.
We specifically look at top jets as they are the ones which have the most complex substructure and in our experience have been the hardest to model.
In~\cref{fig:hlvs-30}, we see that \jeditwo has a much improved $\tau_{32}$ modelling compared to \pcjedi, and has an excellent agreement with Delphes simulation for all other substructure variables.

The generation performance is summarised in~\cref{tab:30_table} for gluon and top jets.
It is observed that MPGAN and \pcjedi already approach the ideal limit for several of the metrics introduced in Ref.~\cite{MPGAN}, thus we focus on metrics where performance gain can be achieved.
Particularly in FPND, \jeditwo manages to significantly bridge that gap and surpasses all methods in all metrics.

\begin{table*}[tp]
    \centering
    \caption{Comparison of generative models on top and gluons with up to 30 constituents. Lower is better.}
    \label{tab:30_table}
    \resizebox{\textwidth}{!}{%
        \begin{tabular}{llccccccc}
\toprule
Jet & Model & FPND & $\mathrm{W}_1^\mathrm{P}$ $(\times 10^{-4})$ & $\mathrm{W}_1^\mathrm{EFP}$ $(\times 10^{-6})$ & $\mathrm{W}_1^\mathrm{M}$ $(\times 10^{-4})$ & W$_1^{\tau_{21}}$ $(\times 10^{-3})$ & W$_1^{\tau_{32}}$ $(\times 10^{-3})$ & W$_1^{D_2}$ $(\times 10^{-1})$ \\
\midrule

\multirow{4}{*}{Top}
& Delphes & $0.01$ & $3.98 \pm 1.27$ & $8.07 \pm 3.51$ & $3.23 \pm 1.07$ & $2.01 \pm 0.74$ & $2.90 \pm 1.59$ & $3.34 \pm 1.03$ \\ \cline{2-9}
& \jeditwo & $\mathbf{0.02}$ & $\mathbf{5.02 \pm 1.59}$ & $\mathbf{11.59 \pm 3.29}$ & $\mathbf{4.27 \pm 1.39}$ & $\mathbf{2.91 \pm 1.09}$ & $\mathbf{5.14 \pm 1.06}$ & $\mathbf{4.75 \pm 1.26}$ \\
& PC-JeDi & $0.15$ & $12.07 \pm 2.01$ & $35.61 \pm 4.92$ & $13.64 \pm 3.21$ & $4.55 \pm 1.16$ & $16.05 \pm 1.31$ & $14.12 \pm 1.48$ \\
& MPGAN & $0.36$ & $21.73 \pm 2.02$ & $12.80 \pm 4.89$ & $6.41 \pm 2.09$ & $6.61 \pm 0.92$ & $17.41 \pm 2.78$ & $11.34 \pm 1.03$ \\
\midrule

\multirow{4}{*}{Gluon}
& Delphes & $0.01$ & $3.54 \pm 1.19$ & $4.07 \pm 1.27$ & $4.39 \pm 1.59$ & $3.79 \pm 1.42$ & $2.26 \pm 0.51$ & $3.72 \pm 1.07$ \\ \cline{2-9}
& \jeditwo & $\mathbf{0.01}$ & $\mathbf{3.66 \pm 1.07}$ & $\mathbf{4.13 \pm 1.61}$ & $\mathbf{4.48 \pm 1.47}$ & $\mathbf{2.89 \pm 0.80}$ & $\mathbf{1.99 \pm 0.51}$ & $\mathbf{4.14 \pm 1.30}$ \\
& PC-JeDi & $0.10$ & $5.83 \pm 1.44$ & $5.68 \pm 1.09$ & $5.66 \pm 1.51$ & $12.48 \pm 0.98$ & $13.32 \pm 0.96$ & $10.70 \pm 2.60$ \\
& MPGAN & $0.13$ & $10.26 \pm 1.51$ & $8.76 \pm 2.44$ & $8.15 \pm 2.10$ & $16.83 \pm 2.08$ & $25.27 \pm 1.29$ & $24.88 \pm 2.91$ \\ \bottomrule

\end{tabular}
    }
\end{table*}

  \subsection{JetNet150}

We extend this work to the 150 constituent dataset and in addition to the original transformer based model we also evaluate CAE networks configured with 1 and 16 global tokens, labelled \jeditwo (CAE-1) and \jeditwo (CAE-16) respectfully.

We find that the LMS sampling method at 100 NFE provides a similarly optimal result for both networks on this dataset.

\begin{figure*}[htpb]
    \centering
        \includegraphics[width=0.33\linewidth]{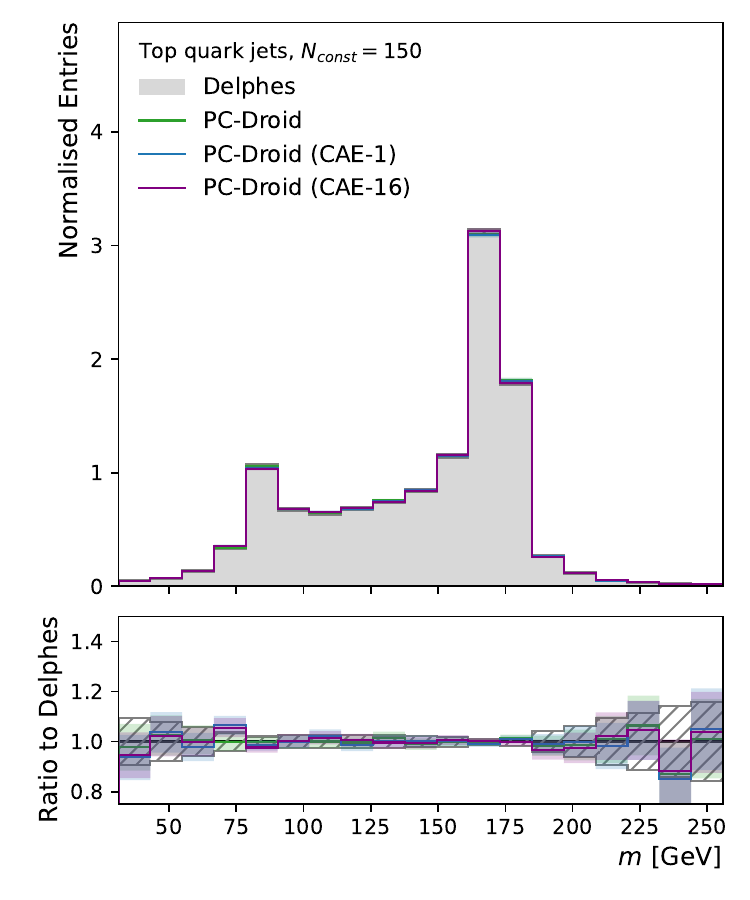}
        \includegraphics[width=0.33\linewidth]{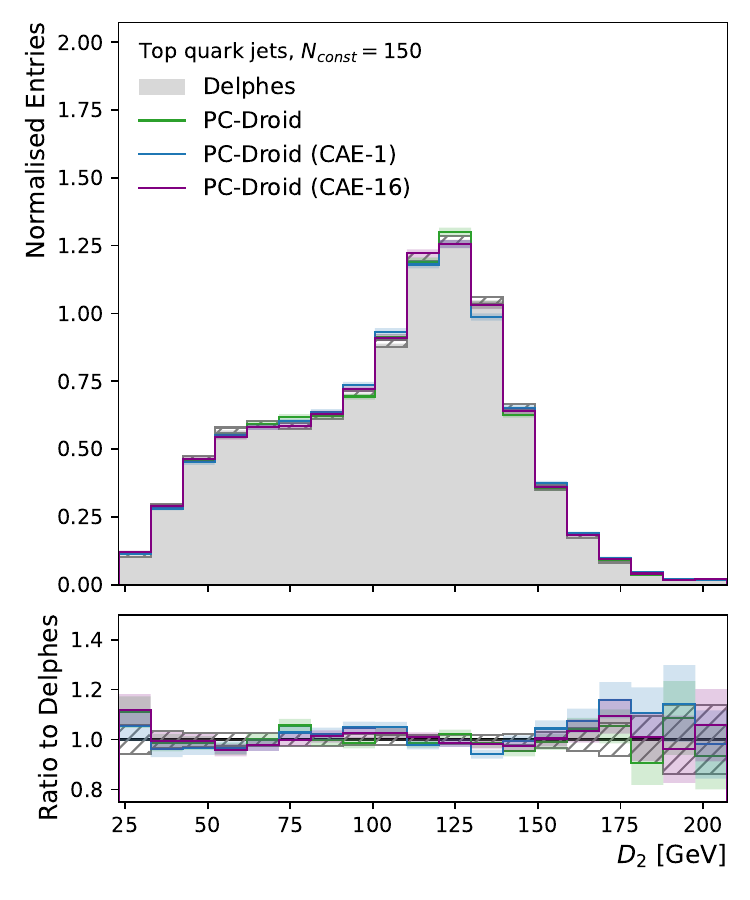} \\
        \includegraphics[width=0.33\linewidth]{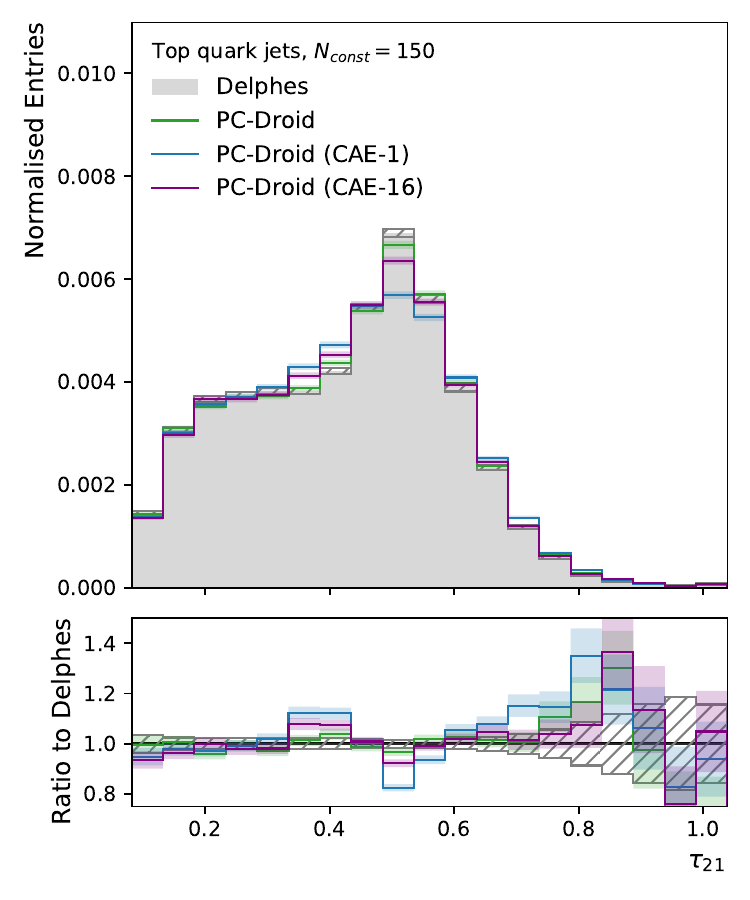}
        \includegraphics[width=0.33\linewidth]{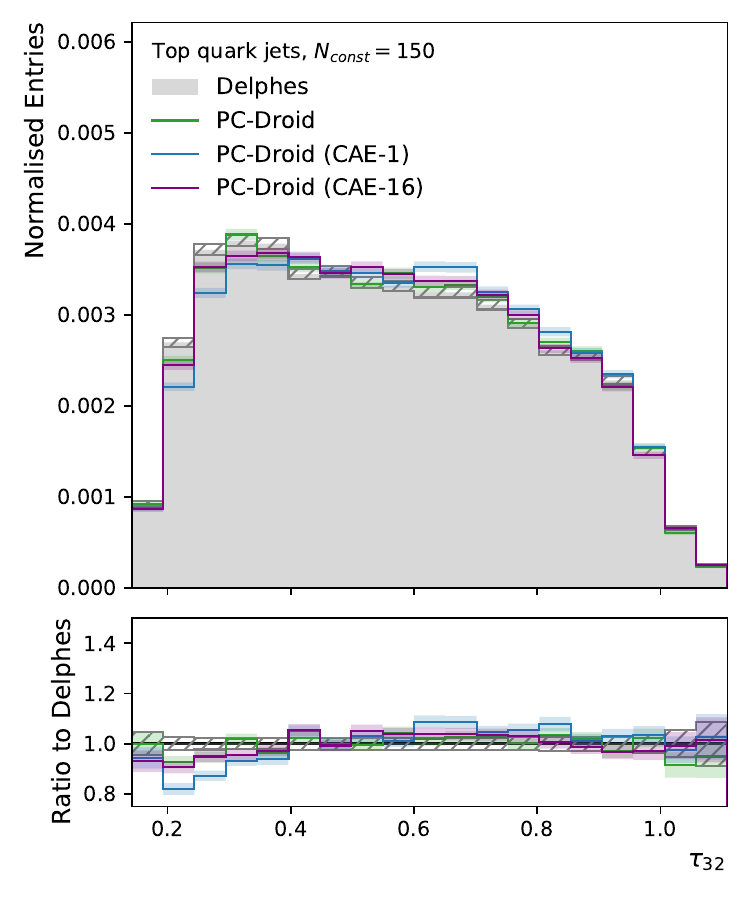}
        \caption{
        Comparison of the standard transformer based \jeditwo and the cross-attention encoder (CAE) variants using the generated mass and substructure marginals of top jets with up to 150 constituents.
        }
    \label{fig:hlvs-150-marginals}
\end{figure*}

\begin{figure*}[htpb]
    \centering
        \includegraphics[width=0.49\linewidth]{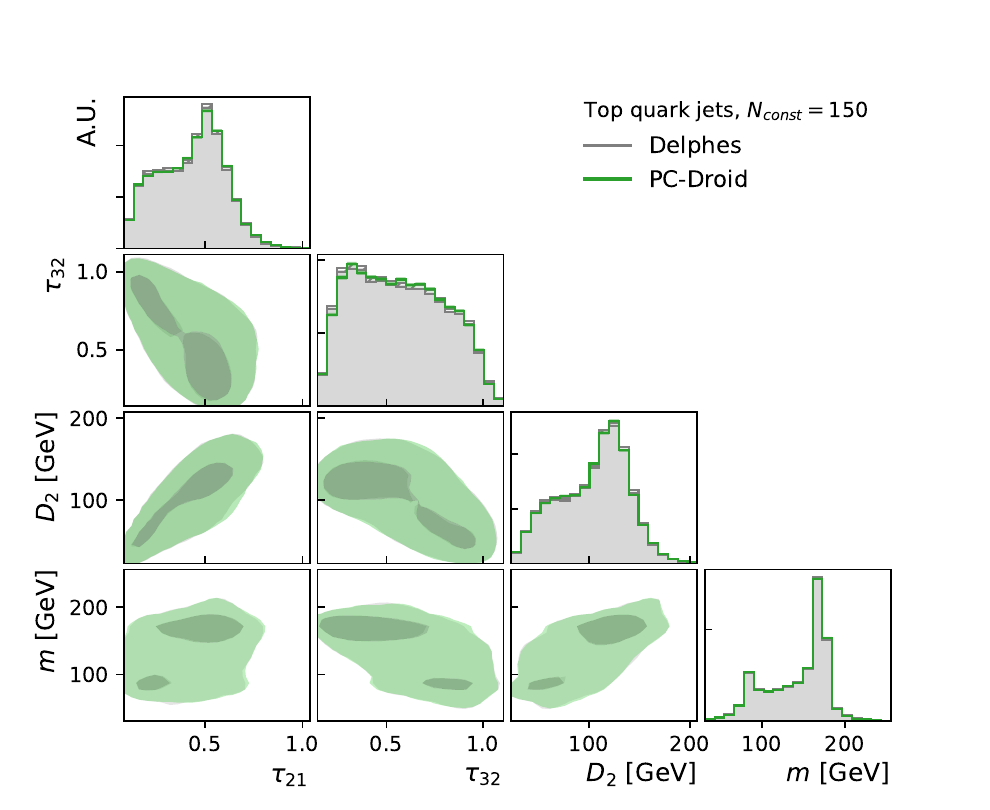}
        \includegraphics[width=0.49\linewidth]{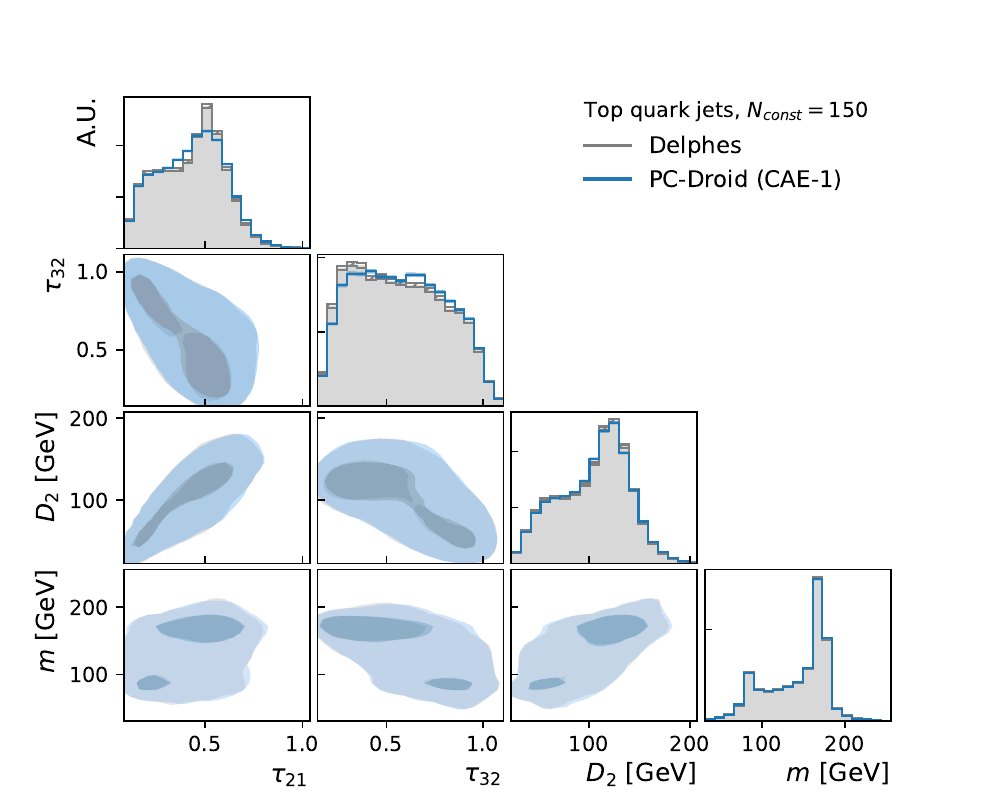}
        \caption{
        Mass and substructure distributions of the generated top jets with up to 150 constituents.
        The diagonal consists of the marginals of the distributions and the off-diagonal elements contain the joint distributions.
        }
    \label{fig:hlvs-150}
\end{figure*}

Substructure distributions are shown in \cref{fig:hlvs-150-marginals}.
Both models do very well in capturing jet mass and $D_2$, but the CAEs show a slight drop in performance when looking at the subjettiness ratios.
This can also be seen in their correlations, as shown in \cref{fig:hlvs-150}.

We provide a quantitative comparison between the performance of the \jeditwo models and EPiC-GAN in \cref{tab:perf-150}.
\jeditwo with the self attention network displays superior performance in all metrics where there is notable separation between models.
EPiC-GAN seems to offer superior performance in modelling the relative mass, but all generative modes are within agreement to Delphes on this metric.

\begin{table*}[tp]
    \centering
    \caption{Comparison of generative models on top and gluon jets with up to 150 constituents. Lower is better. The FPND score is only sensitive to the leading 30 constituents.}
    \label{tab:perf-150}
    \resizebox{\textwidth}{!}{%
        \begin{tabular}{llccccccc}
\toprule
Jet & Model & FPND & $\mathrm{W}_1^\mathrm{P}$ $(\times 10^{-4})$ & $\mathrm{W}_1^\mathrm{EFP}$ $(\times 10^{-6})$ & $\mathrm{W}_1^\mathrm{M}$ $(\times 10^{-4})$ & W$_1^{\tau_{21}}$ $(\times 10^{-3})$ & W$_1^{\tau_{32}}$ $(\times 10^{-3})$ & W$_1^{D_2}$ $(\times 10^{-1})$ \\

\midrule

\multirow{4}{*}{Top}
& Delphes & $0.01$ & $3.03 \pm 0.78$ & $14.77 \pm 5.61$ & $3.96 \pm 0.94$ & $1.78 \pm 0.56$ & $2.78 \pm 1.03$ & $3.82 \pm 1.54$ \\ \cline{2-9}
& \jeditwo & $\mathbf{0.01}$ & $5.45 \pm 1.70$ & $\mathbf{15.41 \pm 5.38}$ & $\mathbf{3.60 \pm 1.20}$ & $\mathbf{2.87 \pm 1.20}$ & $\mathbf{4.70 \pm 1.88}$ & $\mathbf{4.07 \pm 0.85}$ \\
& \jeditwo (CAE-1) & $0.06$ & $\mathbf{4.03 \pm 1.37}$ & $19.93 \pm 6.76$ & $4.28 \pm 1.51$ & $5.21 \pm 0.71$ & $13.01 \pm 2.08$ & $5.89 \pm 1.10$ \\
& \jeditwo (CAE-16) & $0.02$ & $5.40 \pm 1.43$ & $17.17 \pm 5.85$ & $4.63 \pm 1.28$ & $3.22 \pm 0.61$ & $5.94 \pm 1.55$ & $5.37 \pm 2.10$ \\
& EPiC-GAN & $0.95$ & $36.70 \pm 1.80$ & $30.29 \pm 5.01$ & $6.93 \pm 1.49$ & $13.28 \pm 1.80$ & $37.56 \pm 1.92$ & $19.03 \pm 2.40$ \\
\midrule

\multirow{4}{*}{Gluon}
& Delphes & $0.01$ & $3.88 \pm 1.24$ & $10.66 \pm 3.30$ & $6.04 \pm 2.18$ & $2.92 \pm 1.20$ & $1.74 \pm 0.45$ & $4.65 \pm 1.35$ \\ \cline{2-9}
& \jeditwo & $\mathbf{0.01}$ & $3.69 \pm 1.35$ & $10.30 \pm 4.36$ & $5.26 \pm 2.43$ & $\mathbf{3.13 \pm 1.10}$ & $\mathbf{2.24 \pm 0.93}$ & $\mathbf{4.81 \pm 1.27}$ \\
& \jeditwo (CAE-1) & $\mathbf{0.01}$ & $\mathbf{3.38 \pm 1.10}$ & $11.84 \pm 4.15$ & $\mathbf{4.62 \pm 1.29}$ & $4.07 \pm 1.30$ & $2.41 \pm 0.75$ & $5.67 \pm 2.38$ \\
& \jeditwo (CAE-16) & $\mathbf{0.01}$ & $4.37 \pm 1.43$ & $\mathbf{10.09 \pm 2.94}$ & $5.04 \pm 1.77$ & $3.22 \pm 1.06$ & $\mathbf{2.24 \pm 0.72}$ & $5.04 \pm 1.63$ \\
& EPiC-GAN & $0.54$ & $32.22 \pm 1.83$ & $12.72 \pm 4.14$ & $5.00 \pm 1.47$ & $15.49 \pm 2.14$ & $13.32 \pm 1.08$ & $18.61 \pm 1.63$ \\
\bottomrule
\end{tabular}

    }
\end{table*}

\begin{figure}[htpb]
    \centering
    \includegraphics[width=0.8\linewidth]{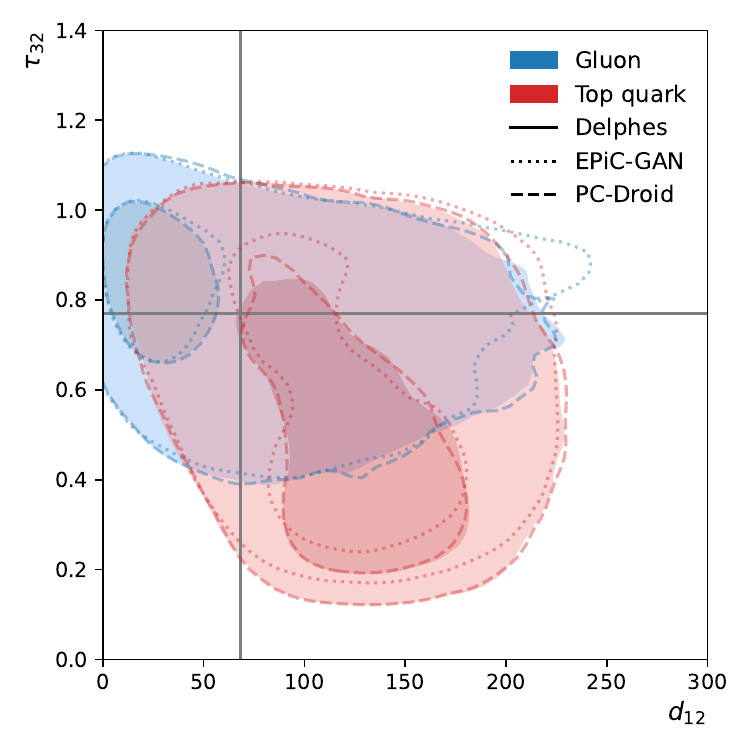}
    \caption{Decision boundaries for the 2D cut based tagger using $d_{12}$ and $\tau_{32}$ for top and gluon jets with up to 150 constituents.}
    \label{fig:dt-150}
\end{figure}

Comparing the performance of models in the context of jet tagging is also useful, as ideally a jet classifier has the same separation power on the generated and reference samples.
The FPND score introduced in Ref~\cite{MPGAN} attempts to capture this, where the jet classifier is a message passing neural network.
Here, we present simpler and interpretable method using a 2D cut based tagger, similar to those used by the ATLAS collaboration \cite{ATLASSubTag0, ATLASSubTag1, ATLASSubTag2}.
We define a simple cut based top/gluon tagger using the $\tau_{32}$ and $d_{12}$ variables.
We do not use the combined jet mass as a discriminating variable as we choose to focus on the substructure of the jets.
The cuts are selected to maximise the signal significance ($S/\sqrt{B}$) at a signal efficiency targeting $70\%$.
The efficiencies for test set and generated jets are summarized in Table~\ref{tab:eff-150}.
Notably, the efficiencies of \jeditwo are very close to those of Delphes, while EPiC-GAN exhibits lower efficiency for both classes of jet.
The decision boundaries are shown in \cref{fig:dt-150}, which also shows that the variables are better modelled by \jeditwo.

\begin{table}[tp]
    \centering
    \caption{Top jet selection efficiency and gluon rejection efficiency (both in \%) for the Delphes simulation and jets generated with \jeditwo and EPiC-GAN. Jets are generated with up to 150 constituents.}
    \label{tab:eff-150}
    \resizebox{0.45\textwidth}{!}{%
        \begin{tabular}{lcc}
\toprule
Model & Top Efficiency & Gluon Rejection \\
\midrule
Delphes           & $70.0 \pm 0.36$ & $83.6 \pm 0.17$ \\ \cline{0-2}
\jeditwo          & $70.0 \pm 0.36$ & $83.7 \pm 0.17$ \\
\jeditwo (CAE-1)  & $69.1 \pm 0.36$ & $83.5 \pm 0.17$ \\
\jeditwo (CAE-16) & $70.3 \pm 0.36$ & $83.8 \pm 0.17$ \\
EPiC-GAN          & $65.1 \pm 0.34$ & $86.5 \pm 0.16$ \\
\bottomrule
\end{tabular}
    }
\end{table}

\subsection{Conditional adherence}

\begin{figure}[htpb]
    \centering
    \includegraphics[width=0.8\linewidth]{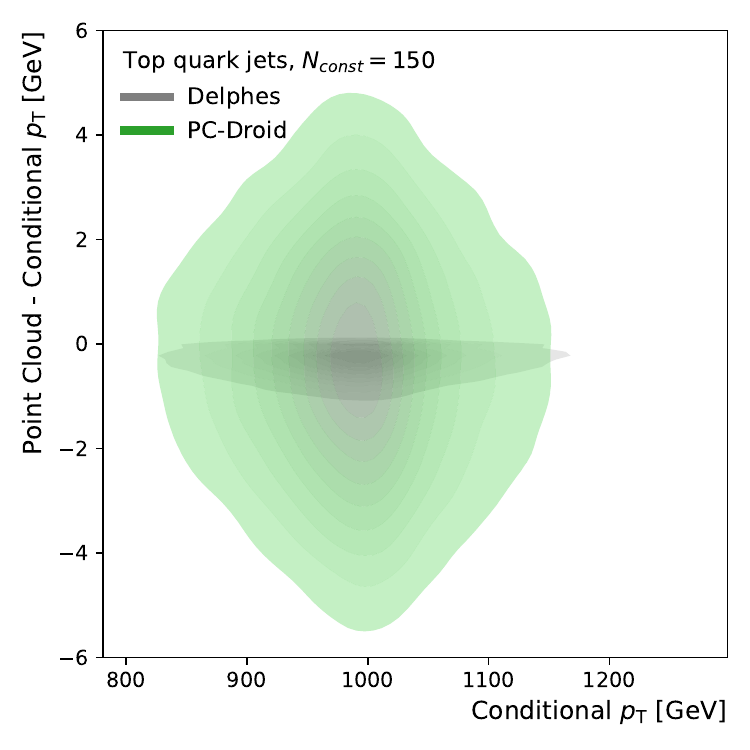}\\
    \includegraphics[width=0.8\linewidth]{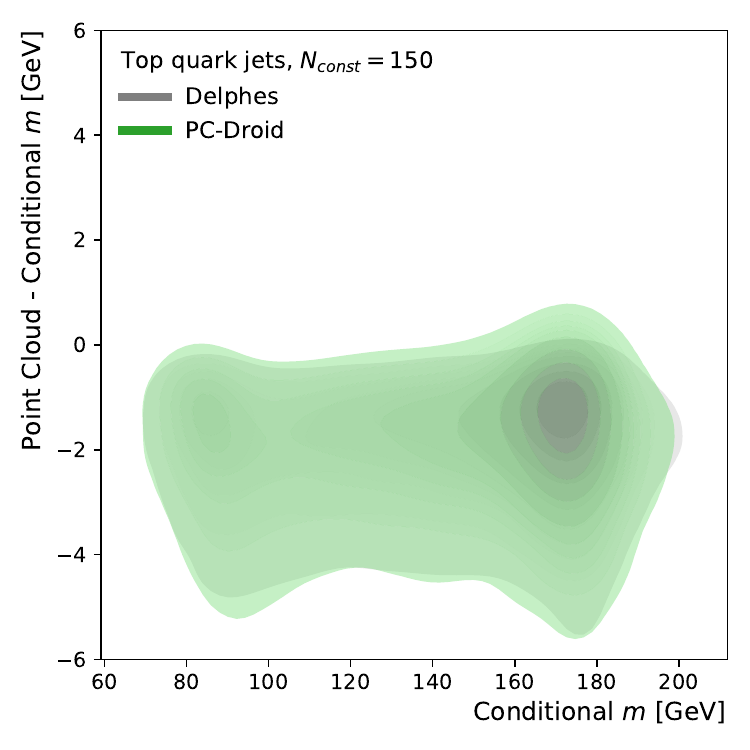}
    \caption{The correlation between the conditional and point cloud kinematics on top jets with up to 150 constituents. The $y$-axis, shows the difference between the two variables.}
    \label{fig:obedience}
\end{figure}

As \jeditwo is a conditional generative model it is of interest to see if the model generates jets with kinematics following the conditioning values. 
To test this we look at the difference between the reconstructed jet kinematics and the target conditions for mass and \pt in \cref{fig:obedience}.

It is important to note that the exact jet kinematics from simulation are used for the \textit{conditional kinematics} whilst the \textit{point cloud kinematics} are recalculated from the preprocessed jet constituents.
These different approaches result in slight discrepancies between the conditional and point cloud variables, even for the Delphes dataset, though it is minor.
Furthermore, the original jets may have had even more than 150 constituents when the kinematics were calculated, which is why the shift in mass and \pt is negative.

For the mass plot in \cref{fig:obedience} we see \jeditwo has a near identical spread compared to Delphes.
For \pt there is slightly more variation as the residual magnitudes are higher, but \jeditwo is within $0.4\%$ of the target.

\subsection{Consistency models}

All diffusion models offer a trade off between the number of iterations and the quality of the final sample.
Consistency models extends this trade off as they are able produce realistic samples with as few as one iteration step, although not with the same fidelity.
When restricted to only a few number of steps, they substantially outperform diffusion models.

In order to improve the generation speed and study its impact on the performance, we train a consistency model using \jeditwo as the teacher.
We use the same training configuration as the original paper \cite{song2023consistency} with one modification.
Instead of using the Heun integration solver to select adjacent points of the ODE, we find DPM2 leads to improved performance.
For generation, we are able to achieve reasonable performance with just a single network pass and observe that performance saturates after five steps.

\begin{figure*}[htpb]
    \centering
        \includegraphics[width=0.49\linewidth]{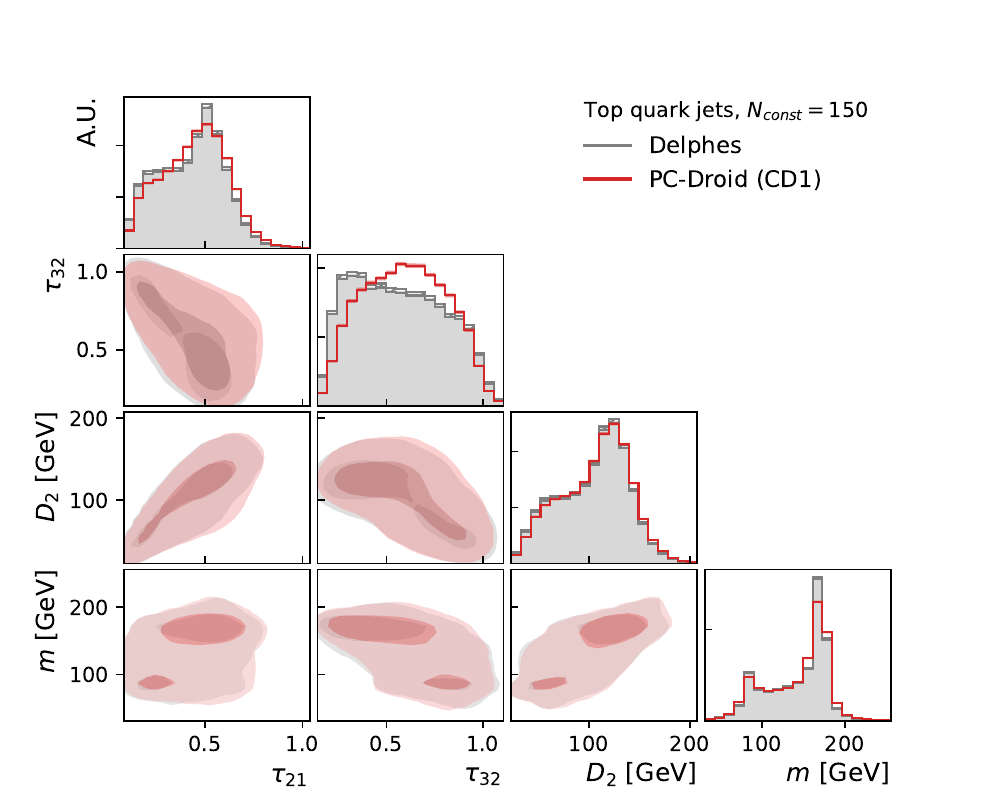}
        \includegraphics[width=0.49\linewidth]{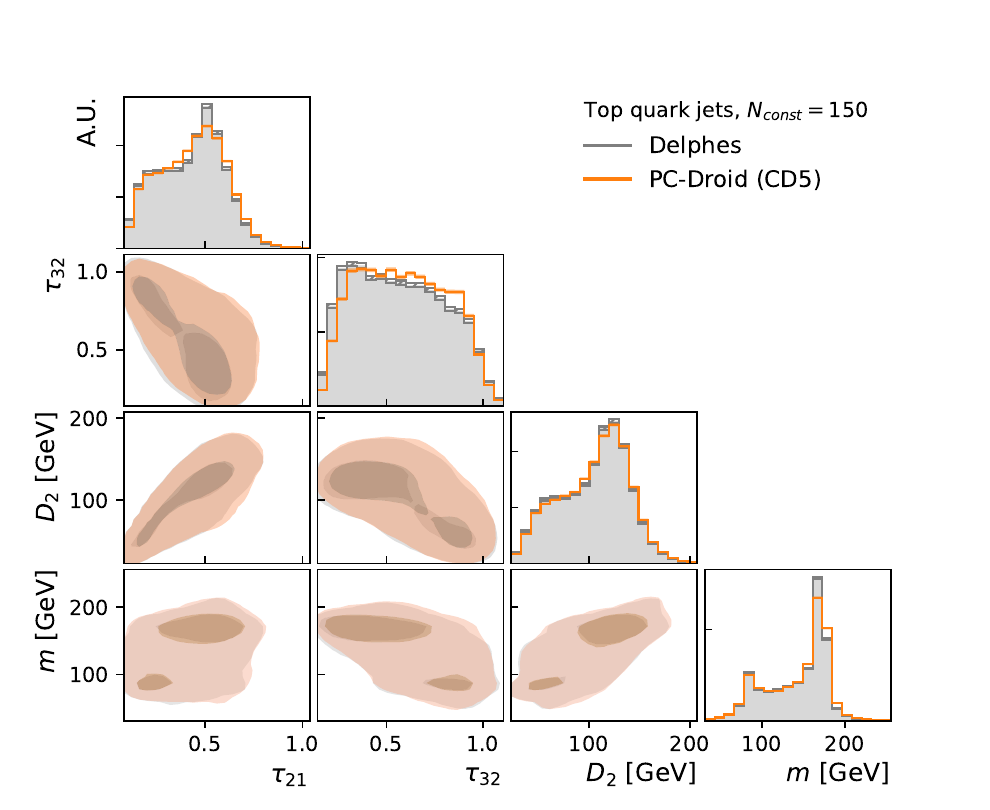}
    \caption{
        Mass and substructure distributions of the generated top jets with up to 150 constituents using the CD models.
        The diagonal consists of the marginals and the off-diagonal elements contain the joint distributions.
        }
    \label{fig:cd-150}
\end{figure*}

The substructure variables for jets produced with the CD model using one-step (CD1) and multi-step (CD5) generation are compared in \cref{fig:cd-150}.
The top jet subjettiness ratios prove again to be the hardest to reproduce, but the one-shot consistency model also demonstrates an overall smear in the mass.

\begin{figure}[htbp]
    \centering
    \includegraphics[width=0.85\linewidth]{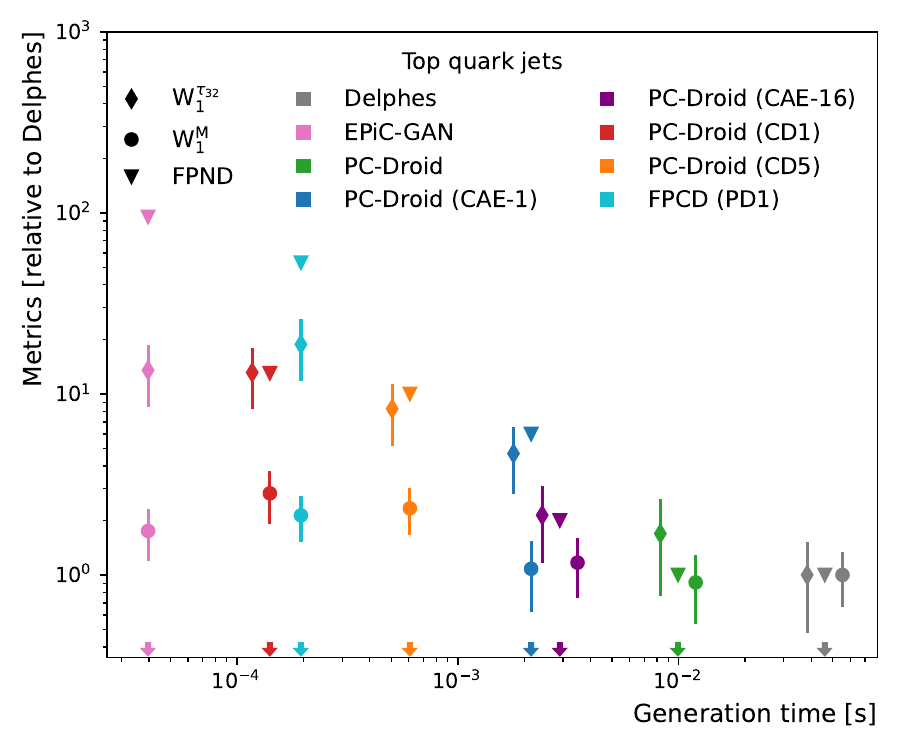}
    \caption{Performance as a ratio to Delphes as a function of the required generation time for a top jet with up to 150 constituents. The time for Delphes is taken from Ref.~\cite{MPGAN}.} 
    \label{fig:timing}
\end{figure}

We compare the quality-generation time trade-off for all models.
\Cref{fig:timing} shows three representative metrics for all models as a function of the generation time for producing top jets with 150 constituents on the same hardware.%
\footnote{All times are calculated from the average of ten runs each generating a batch of 512 jets using an \nvidia GeForce RTX 3080.}
All scores are calculated with respect to the ideal performance, defined by the Delphes dataset.

Our fastest model (CD1) is around three times slower than EPiC-GAN and produces a slight shift in the mass of the jets.
However, the FPND score is an order of magnitude closer to the ideal score, with $\tau_{32}$ comparable between the two.
CD5 is a further five times slower than the one-shot generation, but brings improvements to all metrics.
CAE-1 further improves all metrics, particularly $\mathrm{W}_1^\mathrm{M}$.
Increasing the number of global tokens to 16 results in a 50\% longer generation time but a much improved FPND and $\mathrm{W}_1^{\tau_{32}}$.
The full model is around 250 times slower than EPiC-GAN, but has near ideal performance in all but $\mathrm{W}_1^{\tau_{32}}$.
In comparison to the PD models in Ref.~\cite{FPCD}, our fastest consistency model is just under 30\% faster than FPCD~(PD1), with notably improved performance, particularly in FPND.
When comparing multi-step distillation approaches, CD5 is 2.5 times faster than the eight-step FPCD~(PD8) model. %


  \section{Conclusion}
  In this work we have introduced an updated version of the \pcjedi model for generating jets as particle clouds, called \jeditwo.
An improved diffusion noise scheduler and training procedure, as well as more modern integration solvers, combine to yield state-of-the-art results across a wide range of metrics.
We also consider more jet types and provide a clean and simple method to perform unconditional generation.

We study an additional network architecture to optimise the trade-off between speed and generation quality, and demonstrate the potential of consistency models for one-shot generation.
Even with our fastest models we outperform other competing methods across several of the studied metrics.

Due to its success at jet generation, we expect the findings in this work to be similarly competitive at generating other physical point clouds.
A natural application is for the simulation of particle showers in calorimeters, which has already seen successful and encouraging performance from the application of diffusion models~\cite{CaloScore,CaloClouds}.

  \section*{Acknowledgements}
  The authors would like to acknowledge funding through the SNSF Sinergia grant called ``Robust Deep Density Models for High-Energy Particle Physics and Solar Flare Analysis (RODEM)'' with funding number CRSII$5\_193716$ and the SNSF project grant 200020\_212127 called ``At the two upgrade frontiers: machine learning and the ITk Pixel detector''.
They would also like to acknowledge the funding from the Swiss Government Excellence Scholarships for Foreign Scholars.

  \bibliography{main.bib}

  \clearpage
  \appendix
  \section{Additional Material}
  \subsection{Modelling the jet kinematics}

\begin{figure*}[b!]
    \centering
        \includegraphics[width=0.49\linewidth]{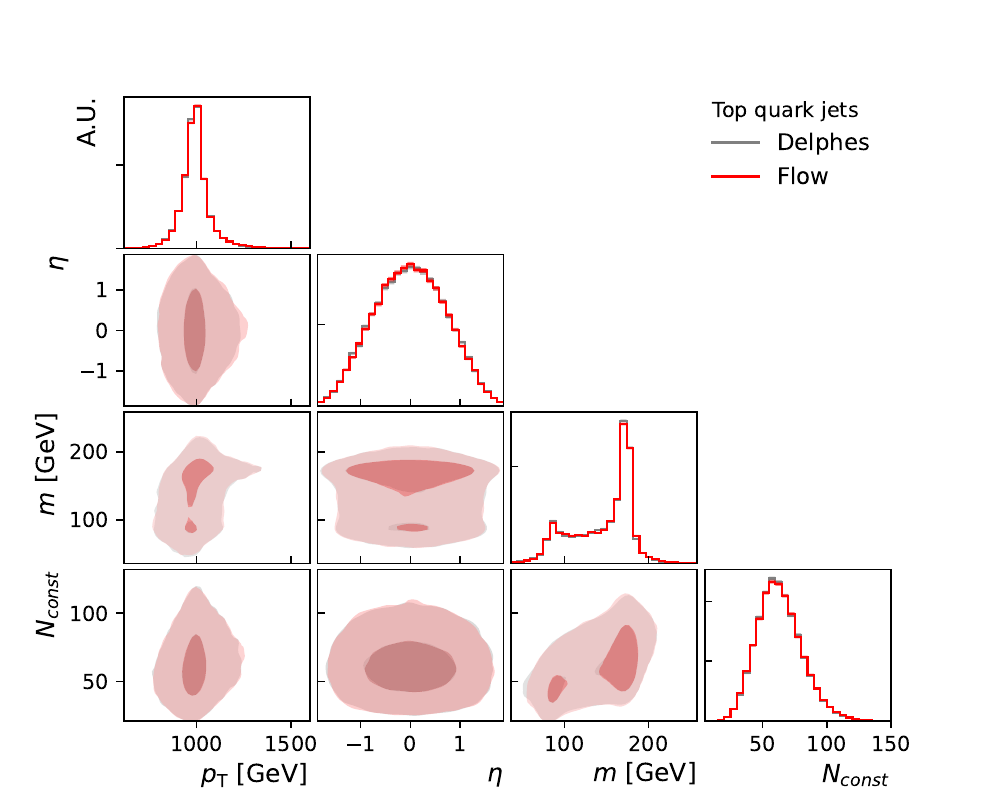}
        \includegraphics[width=0.49\linewidth]{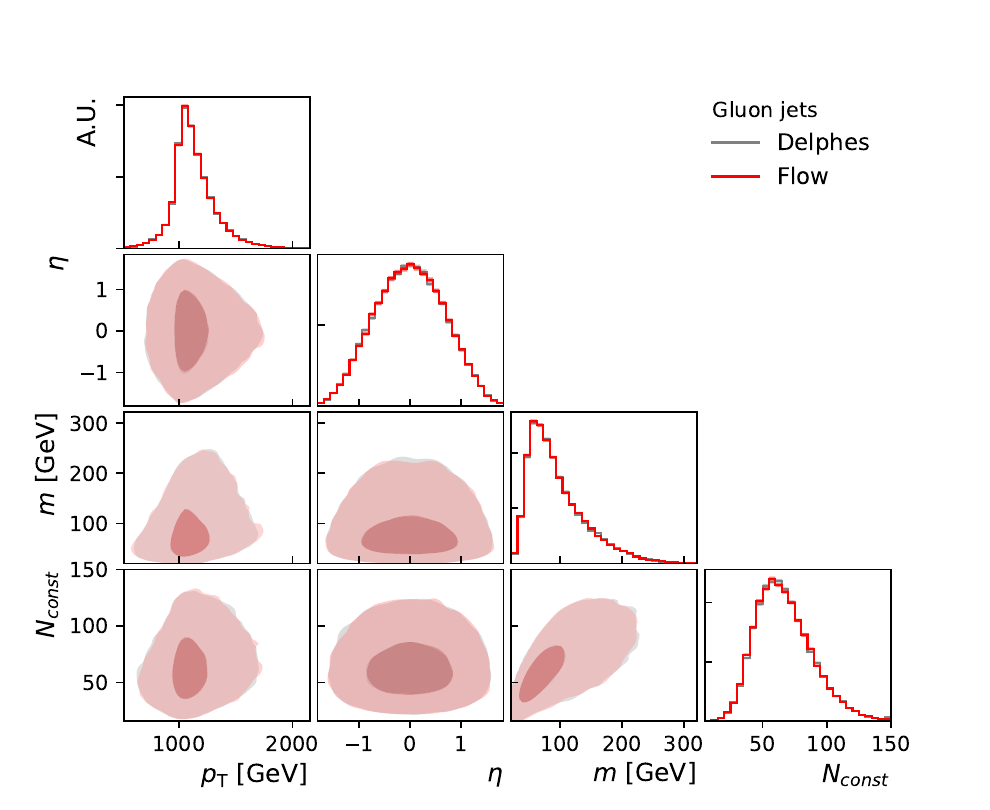}
    \caption{Marginals of the conditioning variables generated with Flow-$\vec{p}$ (red) compared to Delphes (black) for top and gluon jets with up to 150 constituents.
    }
    \label{fig:unconditional-flows}
\end{figure*}

Using normalizing flows we can choose to either sample the full suite of jet kinematics along with $N_{const}$ (Flow-$\vec{p}$) or just $N_{const}$ (Flow-$N$) given the jet kinematics.
Both Flow-$\vec{p}$ and Flow-$N$ are conditional on PID.
It is important to note that these flows are trained independent of \jeditwo as this is a completely orthogonal task.
The performance of Flow-$\vec{p}$ is demonstrated in \cref{fig:unconditional-flows}, where we observe that the marginals and correlations of the conditioning variables are accurately modelled.

We compare the conditional and unconditional models using the full set of metrics in \cref{tab:perf-150-unconditional}.
Here \jeditwo represents the conditional model, where the $N_{const}$ and the jet kinematics come from the test set.
The other two rows represent using either flow to first sample these variables before passing to the diffusion model.
There very little difference in the performance when moving to unconditional generation, and \pcjedi is still notably outperforming other methods.

\begin{table*}[tp]
    \centering
    \caption{Comparison of conditional and unconditional models on jets with up to 150 constituents. Lower is better for all metrics except Cov. The FPND score is only sensitive to the leading 30 constituents.
    }
    \label{tab:perf-150-unconditional}
    \resizebox{\textwidth}{!}{%
        \begin{tabular}{llccccccccc}
\toprule
Jet & Model & FPND & $\mathrm{W}_1^\mathrm{P}$ $(\times 10^{-4})$ & $\mathrm{W}_1^\mathrm{EFP}$ $(\times 10^{-6})$ & $\mathrm{W}_1^\mathrm{M}$ $(\times 10^{-4})$ & W$_1^{\tau_{21}}$ $(\times 10^{-3})$ & W$_1^{\tau_{32}}$ $(\times 10^{-3})$ & W$_1^{D_2}$ $(\times 10^{-1})$ & MMD $(\times 10^{-1})$ & Cov $\uparrow$ \\
\midrule

\multirow{4}{*}{Top}
& Delphes & $0.01$ & $3.03 \pm 0.78$ & $14.77 \pm 5.61$ & $3.96 \pm 0.94$ & $1.78 \pm 0.56$ & $2.78 \pm 1.03$ & $3.82 \pm 1.54$ & $0.48$ & $0.58$ \\ \cline{2-11}
& PC-Droid & $\mathbf{0.01}$ & $5.45 \pm 1.70$ & $15.41 \pm 5.38$ & $\mathbf{3.60 \pm 1.20}$ & $\mathbf{2.87 \pm 1.20}$ & $\mathbf{4.70 \pm 1.88}$ & $\mathbf{4.07 \pm 0.85}$ & $0.49$ & $\mathbf{0.58}$ \\
& PC-Droid (Flow-$\vec{p}$) & $0.02$ & $4.95 \pm 1.34$ & $17.97 \pm 4.44$ & $4.84 \pm 1.18$ & $3.40 \pm 1.29$ & $5.35 \pm 1.33$ & $4.89 \pm 1.05$ & $\mathbf{0.48}$ & $\mathbf{0.58}$ \\
& PC-Droid (Flow-$N$) & $\mathbf{0.01}$ & $\mathbf{4.89 \pm 1.47}$ & $\mathbf{14.14 \pm 3.93}$ & $3.91 \pm 1.35$ & $3.39 \pm 1.28$ & $5.28 \pm 1.33$ & $4.87 \pm 1.02$ & $0.49$ & $\mathbf{0.58}$ \\
\midrule

\multirow{4}{*}{Gluon}
& Delphes & $0.01$ & $3.88 \pm 1.24$ & $10.66 \pm 3.30$ & $6.04 \pm 2.18$ & $2.92 \pm 1.20$ & $1.74 \pm 0.45$ & $4.65 \pm 1.35$ & $0.26$ & $0.57$ \\ \cline{2-11}
& PC-Droid & $\mathbf{0.01}$ & $3.69 \pm 1.35$ & $10.30 \pm 4.36$ & $5.26 \pm 2.43$ & $\mathbf{3.13 \pm 1.10}$ & $2.24 \pm 0.93$ & $4.81 \pm 1.27$ & $\mathbf{0.27}$ & $\mathbf{0.55}$ \\
& PC-Droid (Flow-$\vec{p}$) & $\mathbf{0.01}$ & $\mathbf{3.63 \pm 1.56}$ & $10.63 \pm 2.07$ & $6.10 \pm 3.02$ & $3.43 \pm 1.19$ & $2.24 \pm 0.91$ & $4.71 \pm 1.16$ & $\mathbf{0.27}$ & $\mathbf{0.56}$ \\
& PC-Droid (Flow-$N$) & $0.02$ & $4.01 \pm 1.22$ & $\mathbf{9.74 \pm 2.53}$ & $\mathbf{4.90 \pm 1.54}$ & $3.40 \pm 1.21$ & $\mathbf{2.20 \pm 0.90}$ & $\mathbf{4.71 \pm 1.14}$ & $\mathbf{0.27}$ & $\mathbf{0.55}$ \\
\midrule

\multirow{4}{*}{Quark}
& Delphes & $0.01$ & $4.87 \pm 1.98$ & $7.38 \pm 2.13$ & $4.54 \pm 1.69$ & $2.79 \pm 0.89$ & $1.92 \pm 0.60$ & $4.04 \pm 0.96$ & $0.17$ & $0.54$ \\ \cline{2-11}
& PC-Droid & $\mathbf{0.01}$ & $5.96 \pm 1.76$ & $\mathbf{6.13 \pm 1.91}$ & $4.42 \pm 1.86$ & $3.58 \pm 0.96$ & $\mathbf{1.88 \pm 0.62}$ & $\mathbf{3.78 \pm 1.30}$ & $\mathbf{0.17}$ & $0.55$ \\
& PC-Droid (Flow-$\vec{p}$) & $\mathbf{0.01}$ & $7.18 \pm 2.33$ & $6.46 \pm 1.42$ & $4.83 \pm 1.41$ & $2.95 \pm 1.04$ & $2.26 \pm 0.79$ & $4.62 \pm 1.12$ & $0.18$ & $\mathbf{0.55}$ \\
& PC-Droid (Flow-$N$) & $\mathbf{0.01}$ & $\mathbf{5.52 \pm 1.86}$ & $6.40 \pm 1.82$ & $\mathbf{4.19 \pm 1.07}$ & $\mathbf{2.92 \pm 1.05}$ & $2.26 \pm 0.78$ & $4.61 \pm 1.10$ & $\mathbf{0.17}$ & $0.55$ \\
\midrule

\multirow{4}{*}{W Boson}
& Delphes & $-$ & $3.86 \pm 1.16$ & $1.87 \pm 0.51$ & $1.73 \pm 0.47$ & $2.79 \pm 0.83$ & $3.19 \pm 1.31$ & $4.56 \pm 1.60$ & $0.18$ & $0.56$ \\ \cline{2-11}
& PC-Droid & $\mathbf{-}$ & $\mathbf{3.32 \pm 0.98}$ & $\mathbf{1.49 \pm 0.38}$ & $1.72 \pm 0.75$ & $3.33 \pm 1.12$ & $\mathbf{2.14 \pm 0.61}$ & $\mathbf{3.45 \pm 0.94}$ & $\mathbf{0.19}$ & $0.55$ \\
& PC-Droid (Flow-$\vec{p}$) & $-$ & $3.77 \pm 1.08$ & $2.07 \pm 0.50$ & $2.27 \pm 0.58$ & $\mathbf{3.02 \pm 1.02}$ & $3.27 \pm 1.04$ & $3.91 \pm 1.20$ & $\mathbf{0.19}$ & $\mathbf{0.56}$ \\
& PC-Droid (Flow-$N$) & $-$ & $3.62 \pm 0.98$ & $1.82 \pm 0.44$ & $\mathbf{1.70 \pm 0.58}$ & $3.08 \pm 1.01$ & $3.31 \pm 1.06$ & $3.95 \pm 1.20$ & $\mathbf{0.19}$ & $0.55$ \\
\midrule

\multirow{4}{*}{Z Boson}
& Delphes & $-$ & $4.12 \pm 1.57$ & $2.10 \pm 0.54$ & $2.19 \pm 0.77$ & $3.10 \pm 1.49$ & $2.00 \pm 0.59$ & $3.78 \pm 1.47$ & $0.20$ & $0.57$ \\ \cline{2-11}
& PC-Droid & $\mathbf{-}$ & $3.90 \pm 1.32$ & $2.01 \pm 0.67$ & $1.95 \pm 0.52$ & $\mathbf{2.71 \pm 0.99}$ & $\mathbf{2.22 \pm 0.62}$ & $\mathbf{3.00 \pm 0.78}$ & $\mathbf{0.21}$ & $\mathbf{0.56}$ \\
& PC-Droid (Flow-$\vec{p}$) & $-$ & $\mathbf{3.62 \pm 1.07}$ & $1.86 \pm 0.39$ & $1.95 \pm 0.69$ & $3.03 \pm 1.05$ & $2.34 \pm 0.67$ & $4.01 \pm 1.08$ & $\mathbf{0.21}$ & $\mathbf{0.56}$ \\
& PC-Droid (Flow-$N$) & $-$ & $4.46 \pm 1.48$ & $\mathbf{1.81 \pm 0.51}$ & $\mathbf{1.70 \pm 0.66}$ & $3.01 \pm 1.06$ & $2.31 \pm 0.67$ & $3.95 \pm 1.09$ & $\mathbf{0.21}$ & $0.55$ \\
\bottomrule

\end{tabular}

    }
\end{table*}

  \section{Additional tables}

In \cref{tab:perf-30-big,tab:perf-150-big} the performance for all models on all five jet types are summarised.
The impact of the number of tokens in the CAE layers, as well as the choice of softplus or sigmoid for the activation function when using one token, are summarised in \cref{tab:perf-150-caes}.
The timing values to generate top quark jets for all models using the same hardware are summarised in \cref{tab:timings}, with models ordered by decreasing generation time.

\begin{table*}[tp]
    \centering
    \caption{Comparison of all models on jets with up to 30 constituents. Upper limit of performance calculated using the Delphes dataset.}
    \label{tab:perf-30-big}
    \resizebox{\textwidth}{!}{\begin{tabular}{llcccccccccc}
\toprule
Jet  & Model & FPND & $\mathrm{W}_1^\mathrm{P}$ $(\times 10^{-4})$ & $\mathrm{W}_1^\mathrm{EFP}$ $(\times 10^{-6})$ & $\mathrm{W}_1^\mathrm{M}$ $(\times 10^{-4})$ & W$_1^{\tau_{21}}$ $(\times 10^{-3})$ & W$_1^{\tau_{32}}$ $(\times 10^{-3})$ & W$_1^{D_2}$ $(\times 10^{-1})$ & MMD $(\times 10^{-1})$ & Cov $\uparrow$ \\
\midrule

\multirow{9}{*}{Top}
& Delphes & $-$ & $0.01$ & $3.98 \pm 1.27$ & $8.07 \pm 3.51$ & $3.23 \pm 1.07$ & $2.01 \pm 0.74$ & $2.90 \pm 1.59$ & $3.34 \pm 1.03$ & $\mathbf{0.58}$ & $0.58$ \\ \cline{2-12}
& MPGAN & $1$ & $0.36$ & $21.73 \pm 2.02$ & $12.80 \pm 4.89$ & $6.41 \pm 2.09$ & $6.61 \pm 0.92$ & $17.41 \pm 2.78$ & $11.34 \pm 1.03$ & $0.59$ & $0.57$ \\
& EPiC-GAN & $1$ & $0.32$ & $20.16 \pm 1.71$ & $20.56 \pm 4.37$ & $7.41 \pm 1.55$ & $11.90 \pm 1.53$ & $14.48 \pm 1.01$ & $31.05 \pm 3.20$ & $\mathbf{0.58}$ & $0.58$ \\
& PC-JeDi & $200$ & $0.15$ & $12.07 \pm 2.01$ & $35.61 \pm 4.92$ & $13.64 \pm 3.21$ & $4.55 \pm 1.16$ & $16.05 \pm 1.31$ & $14.12 \pm 1.48$ & $0.59$ & $0.58$ \\
& PC-Droid & $100$ & $\mathbf{0.02}$ & $\mathbf{5.02 \pm 1.59}$ & $\mathbf{11.59 \pm 3.29}$ & $\mathbf{4.27 \pm 1.39}$ & $\mathbf{2.91 \pm 1.09}$ & $\mathbf{5.14 \pm 1.06}$ & $\mathbf{4.75 \pm 1.26}$ & $0.59$ & $\mathbf{0.59}$ \\
& PC-Droid (CD1) & $1$ & $0.12$ & $6.32 \pm 1.71$ & $27.06 \pm 5.26$ & $5.95 \pm 1.29$ & $15.43 \pm 1.27$ & $34.66 \pm 1.57$ & $25.08 \pm 1.80$ & $0.59$ & $\mathbf{0.59}$ \\
& PC-Droid (CD5) & $5$ & $0.10$ & $5.75 \pm 1.83$ & $20.95 \pm 3.28$ & $5.92 \pm 0.79$ & $9.14 \pm 1.18$ & $26.95 \pm 2.33$ & $16.63 \pm 2.60$ & $0.59$ & $0.58$ \\
& FPCD & $512$ & $0.16$ & $5.90 \pm 1.26$ & $13.06 \pm 2.71$ & $5.50 \pm 0.99$ & $5.71 \pm 0.79$ & $10.72 \pm 1.22$ & $7.63 \pm 0.99$ & $\mathbf{0.58}$ & $0.57$ \\
& FPCD (PD1) & $1$ & $0.53$ & $7.26 \pm 1.68$ & $22.72 \pm 2.71$ & $10.68 \pm 0.89$ & $19.19 \pm 1.10$ & $48.63 \pm 1.68$ & $17.80 \pm 2.74$ & $0.59$ & $0.57$ \\
\midrule

\multirow{9}{*}{Gluon}
& Delphes & $-$ & $\mathbf{0.01}$ & $3.54 \pm 1.19$ & $4.07 \pm 1.27$ & $4.39 \pm 1.59$ & $3.79 \pm 1.42$ & $2.26 \pm 0.51$ & $3.72 \pm 1.07$ & $0.31$ & $0.56$ \\ \cline{2-12}
& MPGAN & $1$ & $0.13$ & $10.26 \pm 1.51$ & $8.76 \pm 2.44$ & $8.15 \pm 2.10$ & $16.83 \pm 2.08$ & $25.27 \pm 1.29$ & $24.88 \pm 2.91$ & $0.31$ & $0.54$ \\
& EPiC-GAN & $1$ & $1.10$ & $16.51 \pm 1.98$ & $5.48 \pm 1.33$ & $5.09 \pm 2.06$ & $21.52 \pm 2.54$ & $13.80 \pm 1.10$ & $22.33 \pm 1.96$ & $\mathbf{0.30}$ & $0.51$ \\
& PC-JeDi & $200$ & $0.10$ & $5.83 \pm 1.44$ & $5.68 \pm 1.09$ & $5.66 \pm 1.51$ & $12.48 \pm 0.98$ & $13.32 \pm 0.96$ & $10.70 \pm 2.60$ & $0.31$ & $\mathbf{0.55}$ \\
& PC-Droid & $100$ & $\mathbf{0.01}$ & $\mathbf{3.66 \pm 1.07}$ & $\mathbf{4.13 \pm 1.61}$ & $\mathbf{4.48 \pm 1.47}$ & $\mathbf{2.89 \pm 0.80}$ & $\mathbf{1.99 \pm 0.51}$ & $4.14 \pm 1.30$ & $\mathbf{0.30}$ & $\mathbf{0.55}$ \\
& PC-Droid (CD1) & $1$ & $0.15$ & $4.21 \pm 1.02$ & $5.94 \pm 2.00$ & $5.16 \pm 2.13$ & $14.91 \pm 1.01$ & $13.22 \pm 0.82$ & $11.98 \pm 3.10$ & $\mathbf{0.30}$ & $\mathbf{0.55}$ \\
& PC-Droid (CD5) & $5$ & $0.17$ & $5.29 \pm 1.61$ & $5.81 \pm 1.46$ & $7.27 \pm 3.14$ & $11.35 \pm 2.58$ & $9.82 \pm 1.04$ & $11.56 \pm 2.22$ & $0.31$ & $\mathbf{0.55}$ \\
& FPCD & $512$ & $0.08$ & $7.75 \pm 1.51$ & $5.75 \pm 1.64$ & $5.78 \pm 2.38$ & $4.68 \pm 1.16$ & $2.68 \pm 0.62$ & $\mathbf{3.85 \pm 1.22}$ & $\mathbf{0.30}$ & $\mathbf{0.55}$ \\
& FPCD (PD1) & $1$ & $0.11$ & $7.70 \pm 1.41$ & $7.86 \pm 1.86$ & $7.05 \pm 1.42$ & $14.83 \pm 1.89$ & $12.32 \pm 0.82$ & $9.17 \pm 3.31$ & $\mathbf{0.30}$ & $0.54$ \\
\midrule

\multirow{8}{*}{Quark}
& Delphes & $-$ & $0.01$ & $4.58 \pm 1.68$ & $3.14 \pm 1.32$ & $3.73 \pm 1.10$ & $3.23 \pm 1.17$ & $1.92 \pm 0.40$ & $4.22 \pm 1.74$ & $\mathbf{0.20}$ & $0.55$ \\ \cline{2-12}
& MPGAN & $1$ & $0.37$ & $50.22 \pm 2.05$ & $6.91 \pm 1.41$ & $6.62 \pm 1.54$ & $9.53 \pm 0.45$ & $21.62 \pm 0.85$ & $21.55 \pm 1.20$ & $0.21$ & $0.50$ \\
& EPiC-GAN & $1$ & $0.47$ & $39.90 \pm 2.25$ & $7.57 \pm 1.53$ & $4.97 \pm 1.28$ & $18.90 \pm 2.24$ & $17.72 \pm 1.07$ & $37.23 \pm 2.64$ & $0.21$ & $0.49$ \\
& PC-Droid & $100$ & $\mathbf{0.02}$ & $4.84 \pm 1.59$ & $\mathbf{3.29 \pm 1.02}$ & $\mathbf{3.87 \pm 1.55}$ & $5.56 \pm 0.93$ & $\mathbf{2.82 \pm 0.65}$ & $\mathbf{4.24 \pm 1.17}$ & $\mathbf{0.20}$ & $\mathbf{0.53}$ \\
& PC-Droid (CD1) & $1$ & $0.29$ & $5.05 \pm 1.51$ & $4.45 \pm 1.52$ & $4.89 \pm 2.46$ & $15.34 \pm 1.80$ & $9.89 \pm 0.84$ & $6.81 \pm 1.97$ & $0.21$ & $0.48$ \\
& PC-Droid (CD5) & $5$ & $0.12$ & $5.96 \pm 1.55$ & $5.55 \pm 2.03$ & $5.65 \pm 2.27$ & $15.30 \pm 2.31$ & $7.47 \pm 0.77$ & $6.20 \pm 2.11$ & $0.21$ & $0.48$ \\
& FPCD & $512$ & $0.05$ & $4.76 \pm 1.71$ & $5.34 \pm 1.59$ & $5.41 \pm 1.47$ & $\mathbf{3.44 \pm 0.82}$ & $4.84 \pm 0.93$ & $6.55 \pm 1.00$ & $0.21$ & $0.49$ \\
& FPCD (PD1) & $1$ & $0.08$ & $\mathbf{4.63 \pm 1.33}$ & $5.90 \pm 1.25$ & $7.03 \pm 1.10$ & $16.33 \pm 1.98$ & $10.81 \pm 0.92$ & $7.66 \pm 2.15$ & $0.22$ & $0.48$ \\
\midrule

\multirow{6}{*}{W Boson}
& Delphes & $-$ & $-$ & $3.70 \pm 0.96$ & $1.17 \pm 0.34$ & $2.05 \pm 0.63$ & $3.12 \pm 1.49$ & $3.04 \pm 1.44$ & $4.16 \pm 1.52$ & $\mathbf{0.23}$ & $0.58$ \\ \cline{2-12}
& PC-Droid & $100$ & $-$ & $\mathbf{3.61 \pm 1.09}$ & $\mathbf{1.02 \pm 0.23}$ & $\mathbf{1.89 \pm 0.83}$ & $3.73 \pm 1.22$ & $3.63 \pm 1.71$ & $4.47 \pm 2.07$ & $\mathbf{0.23}$ & $\mathbf{0.57}$ \\
& PC-Droid (CD1) & $1$ & $-$ & $4.58 \pm 1.08$ & $3.84 \pm 0.44$ & $7.25 \pm 0.41$ & $19.08 \pm 1.58$ & $9.64 \pm 0.71$ & $15.75 \pm 1.64$ & $0.26$ & $0.56$ \\
& PC-Droid (CD5) & $5$ & $-$ & $4.82 \pm 1.03$ & $3.03 \pm 0.30$ & $5.22 \pm 0.47$ & $7.01 \pm 1.83$ & $12.78 \pm 1.34$ & $9.65 \pm 1.66$ & $0.25$ & $\mathbf{0.57}$ \\
& FPCD & $512$ & $-$ & $4.07 \pm 1.24$ & $1.55 \pm 0.44$ & $2.41 \pm 0.53$ & $\mathbf{2.77 \pm 0.82}$ & $\mathbf{2.71 \pm 0.55}$ & $\mathbf{3.22 \pm 1.06}$ & $0.24$ & $\mathbf{0.57}$ \\
& FPCD (PD1) & $1$ & $-$ & $4.68 \pm 0.99$ & $3.51 \pm 0.32$ & $8.33 \pm 0.75$ & $18.78 \pm 1.69$ & $9.53 \pm 0.98$ & $17.07 \pm 1.99$ & $0.25$ & $\mathbf{0.57}$ \\
\midrule

\multirow{6}{*}{Z Boson}
& Delphes & $-$ & $-$ & $3.90 \pm 1.09$ & $1.36 \pm 0.40$ & $2.05 \pm 0.71$ & $2.99 \pm 0.95$ & $2.18 \pm 0.64$ & $3.47 \pm 1.10$ & $\mathbf{0.26}$ & $\mathbf{0.57}$ \\ \cline{2-12}
& PC-Droid & $100$ & $-$ & $4.48 \pm 1.26$ & $\mathbf{1.68 \pm 0.52}$ & $\mathbf{1.95 \pm 0.45}$ & $\mathbf{3.35 \pm 1.11}$ & $3.54 \pm 1.24$ & $3.71 \pm 0.83$ & $\mathbf{0.26}$ & $\mathbf{0.57}$ \\
& PC-Droid (CD1) & $1$ & $-$ & $5.33 \pm 1.40$ & $4.84 \pm 0.45$ & $8.11 \pm 1.16$ & $18.60 \pm 1.65$ & $9.30 \pm 0.79$ & $17.43 \pm 1.98$ & $0.29$ & $0.56$ \\
& PC-Droid (CD5) & $5$ & $-$ & $4.48 \pm 1.03$ & $3.94 \pm 0.48$ & $6.70 \pm 1.17$ & $8.78 \pm 1.87$ & $13.60 \pm 1.27$ & $9.38 \pm 1.20$ & $0.28$ & $\mathbf{0.57}$ \\
& FPCD & $512$ & $-$ & $\mathbf{3.91 \pm 0.89}$ & $1.88 \pm 0.44$ & $2.51 \pm 0.63$ & $3.82 \pm 1.25$ & $\mathbf{2.99 \pm 0.68}$ & $\mathbf{3.34 \pm 0.75}$ & $0.27$ & $\mathbf{0.57}$ \\
& FPCD (PD1) & $1$ & $-$ & $4.28 \pm 1.12$ & $4.82 \pm 0.45$ & $8.93 \pm 0.77$ & $20.40 \pm 1.93$ & $10.81 \pm 1.10$ & $21.13 \pm 1.44$ & $0.28$ & $\mathbf{0.57}$ \\
\bottomrule

\end{tabular}
}
\end{table*}

\begin{table*}[tp]
    \centering
    \caption{Comparison of all models on jets with up to 150 constituents. Upper limit of performance calculated using the Delphes dataset.}
    \label{tab:perf-150-big}
    \resizebox{\textwidth}{!}{\begin{tabular}{llcccccccccc}
\toprule
Jet & Model & NFE & FPND & $\mathrm{W}_1^\mathrm{P}$ $(\times 10^{-4})$ & $\mathrm{W}_1^\mathrm{EFP}$ $(\times 10^{-6})$ & $\mathrm{W}_1^\mathrm{M}$ $(\times 10^{-4})$ & W$_1^{\tau_{21}}$ $(\times 10^{-3})$ & W$_1^{\tau_{32}}$ $(\times 10^{-3})$ & W$_1^{D_2}$ $(\times 10^{-1})$ & MMD $(\times 10^{-1})$ & Cov $\uparrow$ \\
\midrule

\multirow{10}{*}{Top}
& Delphes & $-$ & $\mathbf{0.01}$ & $3.03 \pm 0.78$ & $14.77 \pm 5.61$ & $3.96 \pm 0.94$ & $1.78 \pm 0.56$ & $2.78 \pm 1.03$ & $3.82 \pm 1.54$ & $0.48$ & $0.58$ \\ \cline{2-12}
& EPiC-GAN & $1$ & $0.95$ & $36.70 \pm 1.80$ & $30.29 \pm 5.01$ & $6.93 \pm 1.49$ & $13.28 \pm 1.80$ & $37.56 \pm 1.92$ & $19.03 \pm 2.40$ & $0.58$ & $0.57$ \\
& PC-Droid & $100$ & $\mathbf{0.01}$ & $5.45 \pm 1.70$ & $\mathbf{15.41 \pm 5.38}$ & $\mathbf{3.60 \pm 1.20}$ & $\mathbf{2.87 \pm 1.20}$ & $\mathbf{4.70 \pm 1.88}$ & $\mathbf{4.07 \pm 0.85}$ & $\mathbf{0.49}$ & $0.58$ \\
& PC-Droid (CAE-1) & 100 & $0.06$ & $\mathbf{4.03 \pm 1.37}$ & $19.93 \pm 6.76$ & $4.28 \pm 1.51$ & $5.21 \pm 0.71$ & $13.01 \pm 2.08$ & $5.89 \pm 1.10$ & $\mathbf{0.49}$ & $0.57$ \\
& PC-Droid (CAE-16) & $100$ & $0.02$ & $5.40 \pm 1.43$ & $17.17 \pm 5.85$ & $4.63 \pm 1.28$ & $3.22 \pm 0.61$ & $5.94 \pm 1.55$ & $5.37 \pm 2.10$ & $\mathbf{0.49}$ & $\mathbf{0.59}$ \\
& PC-Droid (CD1) & $1$ & $0.13$ & $6.62 \pm 1.49$ & $55.49 \pm 6.99$ & $11.18 \pm 2.44$ & $19.04 \pm 1.61$ & $36.48 \pm 1.67$ & $18.60 \pm 3.00$ & $0.57$ & $\mathbf{0.59}$ \\
& PC-Droid (CD5) & $5$ & $0.10$ & $6.78 \pm 1.76$ & $54.25 \pm 5.27$ & $9.24 \pm 1.53$ & $7.53 \pm 0.79$ & $22.98 \pm 1.58$ & $10.46 \pm 2.84$ & $0.55$ & $0.58$ \\
& FPCD & $512$ & $0.15$ & $4.36 \pm 1.28$ & $26.11 \pm 6.30$ & $6.44 \pm 2.57$ & $6.30 \pm 0.81$ & $9.91 \pm 1.39$ & $8.59 \pm 1.53$ & $0.50$ & $\mathbf{0.59}$ \\
& FPCD (PD1) & $1$ & $0.53$ & $4.80 \pm 1.31$ & $38.22 \pm 3.65$ & $8.44 \pm 1.30$ & $19.67 \pm 1.43$ & $52.16 \pm 2.12$ & $21.38 \pm 2.28$ & $0.53$ & $0.58$ \\

\midrule

\multirow{10}{*}{Gluon}
& Delphes & $-$ & $\mathbf{0.01}$ & $3.88 \pm 1.24$ & $10.66 \pm 3.30$ & $6.04 \pm 2.18$ & $2.92 \pm 1.20$ & $1.74 \pm 0.45$ & $4.65 \pm 1.35$ & $0.26$ & $\mathbf{0.57}$ \\ \cline{2-12}
& EPiC-GAN & $1$ & $0.54$ & $32.22 \pm 1.83$ & $12.72 \pm 4.14$ & $5.00 \pm 1.47$ & $15.49 \pm 2.14$ & $13.32 \pm 1.08$ & $18.61 \pm 1.63$ & $0.32$ & $0.53$ \\
& PC-Droid & $100$ & $\mathbf{0.01}$ & $3.69 \pm 1.35$ & $10.30 \pm 4.36$ & $5.26 \pm 2.43$ & $\mathbf{3.13 \pm 1.10}$ & $\mathbf{2.24 \pm 0.93}$ & $\mathbf{4.81 \pm 1.27}$ & $\mathbf{0.27}$ & $0.55$ \\
& PC-Droid (CAE-1) & 100 & $\mathbf{0.01}$ & $\mathbf{3.38 \pm 1.10}$ & $11.84 \pm 4.15$ & $\mathbf{4.62 \pm 1.29}$ & $4.07 \pm 1.30$ & $2.41 \pm 0.75$ & $5.67 \pm 2.38$ & $\mathbf{0.27}$ & $\mathbf{0.56}$ \\
& PC-Droid (CAE-16) & $100$ & $\mathbf{0.01}$ & $4.37 \pm 1.43$ & $\mathbf{10.09 \pm 2.94}$ & $5.04 \pm 1.77$ & $3.22 \pm 1.06$ & $\mathbf{2.24 \pm 0.72}$ & $5.04 \pm 1.63$ & $\mathbf{0.27}$ & $0.56$ \\
& PC-Droid (CD1) & $1$ & $0.10$ & $6.54 \pm 1.81$ & $21.55 \pm 6.13$ & $5.55 \pm 2.76$ & $15.33 \pm 2.58$ & $10.73 \pm 0.80$ & $14.11 \pm 2.93$ & $0.31$ & $0.56$ \\
& PC-Droid (CD5) & $5$ & $0.06$ & $7.05 \pm 1.74$ & $14.91 \pm 5.38$ & $7.09 \pm 2.90$ & $11.11 \pm 2.23$ & $9.04 \pm 0.71$ & $14.24 \pm 2.57$ & $0.31$ & $0.56$ \\
& FPCD & $512$ & $0.03$ & $5.14 \pm 1.45$ & $14.59 \pm 4.37$ & $5.51 \pm 2.53$ & $5.64 \pm 1.85$ & $2.74 \pm 0.66$ & $5.10 \pm 1.55$ & $0.28$ & $\mathbf{0.57}$ \\
& FPCD (PD1) & $1$ & $0.08$ & $5.53 \pm 1.24$ & $18.83 \pm 5.71$ & $6.83 \pm 1.61$ & $19.36 \pm 2.45$ & $10.97 \pm 1.03$ & $5.25 \pm 1.67$ & $0.29$ & $0.55$ \\

\midrule

\multirow{10}{*}{Quark}
& Delphes & $-$ & $\mathbf{0.01}$ & $4.87 \pm 1.98$ & $7.38 \pm 2.13$ & $4.54 \pm 1.69$ & $2.79 \pm 0.89$ & $1.92 \pm 0.60$ & $4.04 \pm 0.96$ & $\mathbf{0.17}$ & $0.54$ \\ \cline{2-12}
& EPiC-GAN & $1$ & $0.17$ & $39.69 \pm 2.67$ & $8.19 \pm 3.06$ & $5.25 \pm 1.78$ & $13.00 \pm 1.85$ & $10.35 \pm 0.89$ & $24.01 \pm 2.20$ & $0.19$ & $0.51$ \\
& PC-Droid & $100$ & $\mathbf{0.01}$ & $5.96 \pm 1.76$ & $6.13 \pm 1.91$ & $4.42 \pm 1.86$ & $3.58 \pm 0.96$ & $\mathbf{1.88 \pm 0.62}$ & $\mathbf{3.78 \pm 1.30}$ & $\mathbf{0.17}$ & $\mathbf{0.55}$ \\
& PC-Droid (CAE-1) & 100 & $\mathbf{0.01}$ & $\mathbf{4.55 \pm 1.03}$ & $6.04 \pm 1.86$ & $\mathbf{3.88 \pm 1.53}$ & $4.02 \pm 1.35$ & $2.07 \pm 0.51$ & $4.14 \pm 1.24$ & $\mathbf{0.17}$ & $\mathbf{0.55}$ \\
& PC-Droid (CAE-16) & $100$ & $0.02$ & $5.93 \pm 2.33$ & $\mathbf{5.54 \pm 1.63}$ & $4.82 \pm 1.56$ & $\mathbf{3.52 \pm 0.91}$ & $2.40 \pm 0.66$ & $5.01 \pm 2.07$ & $\mathbf{0.17}$ & $\mathbf{0.55}$ \\
& PC-Droid (CD1) & $1$ & $0.27$ & $7.74 \pm 2.21$ & $9.90 \pm 2.69$ & $6.50 \pm 1.87$ & $13.56 \pm 1.89$ & $9.22 \pm 0.60$ & $6.30 \pm 1.85$ & $0.21$ & $0.54$ \\
& PC-Droid (CD5) & $5$ & $0.15$ & $9.87 \pm 2.07$ & $7.87 \pm 2.19$ & $4.14 \pm 1.54$ & $12.32 \pm 1.65$ & $6.41 \pm 0.71$ & $10.25 \pm 2.08$ & $0.21$ & $0.54$ \\
& FPCD & $512$ & $0.03$ & $5.32 \pm 1.86$ & $5.99 \pm 1.81$ & $7.28 \pm 2.12$ & $3.71 \pm 1.10$ & $3.36 \pm 0.75$ & $7.71 \pm 1.42$ & $0.19$ & $\mathbf{0.55}$ \\
& FPCD (PD1) & $1$ & $0.09$ & $6.39 \pm 1.94$ & $7.24 \pm 2.14$ & $7.43 \pm 2.84$ & $17.58 \pm 1.75$ & $9.40 \pm 0.98$ & $7.89 \pm 2.13$ & $0.20$ & $0.54$ \\

\midrule

\multirow{9}{*}{W Boson}
& Delphes & $-$ & $-$ & $3.86 \pm 1.16$ & $1.87 \pm 0.51$ & $1.73 \pm 0.47$ & $2.79 \pm 0.83$ & $3.19 \pm 1.31$ & $4.56 \pm 1.60$ & $0.18$ & $0.56$ \\ \cline{2-12}
& PC-Droid & $100$ & $-$ & $\mathbf{3.32 \pm 0.98}$ & $\mathbf{1.49 \pm 0.38}$ & $\mathbf{1.72 \pm 0.75}$ & $3.33 \pm 1.12$ & $\mathbf{2.14 \pm 0.61}$ & $\mathbf{3.45 \pm 0.94}$ & $\mathbf{0.19}$ & $0.55$ \\
& PC-Droid (CAE-1) & 100 & $-$ & $3.79 \pm 1.12$ & $1.58 \pm 0.48$ & $2.18 \pm 0.51$ & $3.00 \pm 1.08$ & $3.59 \pm 1.27$ & $3.62 \pm 1.43$ & $\mathbf{0.19}$ & $0.55$ \\
& PC-Droid (CAE-16) & $100$ & $-$ & $3.93 \pm 1.03$ & $1.82 \pm 0.49$ & $2.04 \pm 0.56$ & $\mathbf{2.81 \pm 0.83}$ & $2.62 \pm 1.09$ & $3.69 \pm 0.94$ & $\mathbf{0.19}$ & $0.56$ \\
& PC-Droid (CD1) & $1$ & $-$ & $5.06 \pm 1.16$ & $6.47 \pm 0.89$ & $12.80 \pm 0.93$ & $26.98 \pm 1.57$ & $10.64 \pm 0.73$ & $21.07 \pm 2.23$ & $0.25$ & $0.56$ \\
& PC-Droid (CD5) & $5$ & $-$ & $6.25 \pm 1.07$ & $4.40 \pm 0.62$ & $8.28 \pm 0.43$ & $8.07 \pm 1.48$ & $13.30 \pm 1.44$ & $12.60 \pm 1.05$ & $0.24$ & $0.56$ \\
& FPCD & $512$ & $-$ & $3.53 \pm 1.06$ & $2.02 \pm 0.58$ & $2.97 \pm 0.41$ & $2.86 \pm 0.91$ & $3.91 \pm 0.84$ & $3.67 \pm 1.28$ & $0.22$ & $0.55$ \\
& FPCD (PD1) & $1$ & $-$ & $4.32 \pm 1.40$ & $4.00 \pm 0.57$ & $9.07 \pm 0.51$ & $20.78 \pm 1.48$ & $7.90 \pm 0.45$ & $15.48 \pm 0.99$ & $0.23$ & $\mathbf{0.57}$ \\

\midrule

\multirow{9}{*}{Z Boson}
& Delphes & $-$ & $-$ & $4.12 \pm 1.57$ & $2.10 \pm 0.54$ & $2.19 \pm 0.77$ & $3.10 \pm 1.49$ & $2.00 \pm 0.59$ & $3.78 \pm 1.47$ & $0.20$ & $\mathbf{0.57}$ \\ \cline{2-12}
& PC-Droid & $100$ & $-$ & $3.90 \pm 1.32$ & $\mathbf{2.01 \pm 0.67}$ & $1.95 \pm 0.52$ & $\mathbf{2.71 \pm 0.99}$ & $\mathbf{2.22 \pm 0.62}$ & $\mathbf{3.00 \pm 0.78}$ & $\mathbf{0.21}$ & $0.56$ \\
& PC-Droid (CAE-1) & 100 & $-$ & $\mathbf{3.65 \pm 1.14}$ & $2.04 \pm 0.68$ & $2.24 \pm 0.58$ & $3.43 \pm 1.02$ & $4.16 \pm 1.13$ & $3.51 \pm 0.82$ & $\mathbf{0.21}$ & $\mathbf{0.57}$ \\
& PC-Droid (CAE-16) & $100$ & $-$ & $3.87 \pm 1.15$ & $2.19 \pm 0.53$ & $\mathbf{1.90 \pm 0.73}$ & $3.28 \pm 1.18$ & $2.58 \pm 1.04$ & $5.43 \pm 1.86$ & $\mathbf{0.21}$ & $\mathbf{0.57}$ \\
& PC-Droid (CD1) & $1$ & $-$ & $5.04 \pm 0.94$ & $6.57 \pm 0.73$ & $12.20 \pm 0.53$ & $22.78 \pm 2.07$ & $9.32 \pm 0.67$ & $19.40 \pm 2.78$ & $0.27$ & $0.56$ \\
& PC-Droid (CD5) & $5$ & $-$ & $5.41 \pm 1.21$ & $6.16 \pm 0.74$ & $7.99 \pm 0.45$ & $6.39 \pm 1.28$ & $15.49 \pm 1.23$ & $11.50 \pm 1.89$ & $0.26$ & $0.56$ \\
& FPCD & $512$ & $-$ & $4.66 \pm 1.35$ & $2.59 \pm 0.61$ & $2.97 \pm 0.53$ & $3.90 \pm 1.35$ & $4.26 \pm 1.14$ & $4.12 \pm 1.30$ & $0.23$ & $0.56$ \\
& FPCD (PD1) & $1$ & $-$ & $4.72 \pm 1.26$ & $5.91 \pm 0.75$ & $9.63 \pm 0.93$ & $23.70 \pm 1.76$ & $8.52 \pm 1.08$ & $20.77 \pm 1.98$ & $0.25$ & $\mathbf{0.57}$ \\

\bottomrule

\end{tabular}}
\end{table*}

\begin{table*}[tp]
    \centering
    \caption{Comparisons of different architectures for the cross-attention encoder. We experimented with different numbers of global tokens as well as different activation functions in the cross-attention operation.}
    \label{tab:perf-150-caes}
    \resizebox{\textwidth}{!}{\begin{tabular}{llccccccccc}
\toprule
Jet & Model & FPND & $\mathrm{W}_1^\mathrm{P}$ $(\times 10^{-4})$ & $\mathrm{W}_1^\mathrm{EFP}$ $(\times 10^{-6})$ & $\mathrm{W}_1^\mathrm{M}$ $(\times 10^{-4})$ & W$_1^{\tau_{21}}$ $(\times 10^{-3})$ & W$_1^{\tau_{32}}$ $(\times 10^{-3})$ & W$_1^{D_2}$ $(\times 10^{-1})$ & MMD $(\times 10^{-1})$ & Cov $\uparrow$ \\
\midrule

\multirow{5}{*}{Top}
& Delphes & $0.01$ & $3.03 \pm 0.78$ & $14.77 \pm 5.61$ & $3.96 \pm 0.94$ & $1.78 \pm 0.56$ & $2.78 \pm 1.03$ & $3.82 \pm 1.54$ & $0.48$ & $0.58$ \\ \cline{2-11}
& CAE-1-Sigmoid & $0.06$ & $\mathbf{4.03 \pm 1.37}$ & $19.93 \pm 6.76$ & $4.28 \pm 1.51$ & $5.21 \pm 0.71$ & $13.01 \pm 2.08$ & $5.89 \pm 1.10$ & $\mathbf{0.49}$ & $0.57$ \\
& CAE-1-Softplus & $0.06$ & $5.18 \pm 1.24$ & $15.52 \pm 4.70$ & $4.48 \pm 1.58$ & $5.32 \pm 0.73$ & $13.88 \pm 2.52$ & $6.31 \pm 1.97$ & $\mathbf{0.49}$ & $0.57$ \\
& CAE-4-Softmax & $0.03$ & $5.45 \pm 1.45$ & $\mathbf{14.88 \pm 3.41}$ & $\mathbf{4.22 \pm 1.42}$ & $5.13 \pm 0.51$ & $8.60 \pm 1.94$ & $\mathbf{4.87 \pm 1.18}$ & $\mathbf{0.49}$ & $0.57$ \\
& CAE-16-Softmax & $\mathbf{0.02}$ & $5.40 \pm 1.43$ & $17.17 \pm 5.85$ & $4.63 \pm 1.28$ & $\mathbf{3.22 \pm 0.61}$ & $\mathbf{5.94 \pm 1.55}$ & $5.37 \pm 2.10$ & $\mathbf{0.49}$ & $\mathbf{0.59}$ \\

\midrule

\multirow{5}{*}{Gluon}
& Delphes & $\mathbf{0.01}$ & $3.88 \pm 1.24$ & $10.66 \pm 3.30$ & $6.04 \pm 2.18$ & $2.92 \pm 1.20$ & $1.74 \pm 0.45$ & $4.65 \pm 1.35$ & $0.26$ & $\mathbf{0.57}$ \\ \cline{2-11}
& CAE-1-Sigmoid & $\mathbf{0.01}$ & $\mathbf{3.38 \pm 1.10}$ & $11.84 \pm 4.15$ & $\mathbf{4.62 \pm 1.29}$ & $4.07 \pm 1.30$ & $2.41 \pm 0.75$ & $5.67 \pm 2.38$ & $\mathbf{0.27}$ & $0.56$ \\
& CAE-1-Softplus & $\mathbf{0.01}$ & $4.08 \pm 1.22$ & $11.91 \pm 4.26$ & $4.89 \pm 1.34$ & $3.57 \pm 1.31$ & $3.50 \pm 1.22$ & $\mathbf{4.91 \pm 1.78}$ & $\mathbf{0.27}$ & $0.56$ \\
& CAE-4-Softmax & $\mathbf{0.01}$ & $4.28 \pm 1.34$ & $10.69 \pm 4.16$ & $4.92 \pm 1.54$ & $5.06 \pm 1.47$ & $3.62 \pm 0.87$ & $5.20 \pm 2.43$ & $\mathbf{0.27}$ & $\mathbf{0.57}$ \\
& CAE-16-Softmax & $\mathbf{0.01}$ & $4.37 \pm 1.43$ & $\mathbf{10.09 \pm 2.94}$ & $5.04 \pm 1.77$ & $\mathbf{3.22 \pm 1.06}$ & $\mathbf{2.24 \pm 0.72}$ & $5.04 \pm 1.63$ & $\mathbf{0.27}$ & $0.56$ \\

\bottomrule

\end{tabular}}
\end{table*}

\begin{table*}
    \centering
    \caption{Comparison of required time to generate a top jet with 150 constituents.
    Times are calculated from the average of ten runs each generating a batch of 512 jets using the same hardware with an \nvidia GeForce RTX 3080.
    Models are ordered in increasing time requirement.
For reference, the Delphes simulation required $4.6\times 10^{-2}$~s for each jet~\cite{MPGAN}.
    }
    \label{tab:timings}
    \begin{tabular}{l c}
    \toprule
    Model & Generation time [s]\\
    \midrule
    EPiC-GAN &           $3.96\, (0.00) \times 10^{-5}$   \\
    PC-Droid (CD1) &     $1.42\, (0.01) \times 10^{-5}$   \\
    FPCD (PD1) &         $1.95\, (0.06) \times 10^{-4}$   \\
    PC-Droid (CD5) &     $6.07\, (0.01) \times 10^{-4}$   \\
    FPCD (PD8) &         $1.54\, (0.05) \times 10^{-3}$   \\
    PC-Droid (CAE-1) &   $2.15\, (0.03) \times 10^{-3}$   \\
    PC-Droid (CAE-16) &  $2.93\, (0.02) \times 10^{-3}$   \\
    PC-Droid &           $9.95\, (0.01) \times 10^{-3}$   \\
    FPCD &               $5.99\, (0.00) \times 10^{-2}$   \\
    \bottomrule
\end{tabular}

\end{table*}

  \subsection{Modelling the constituents}

We compare marginals of the leading, fifth leading, and twentieth leading constituents of the generated jets compared to Delphes.
We found that almost all methods were able to perform sufficiently well in this test.

\begin{figure*}[htpb]
    \centering
        \includegraphics[width=0.9\linewidth]{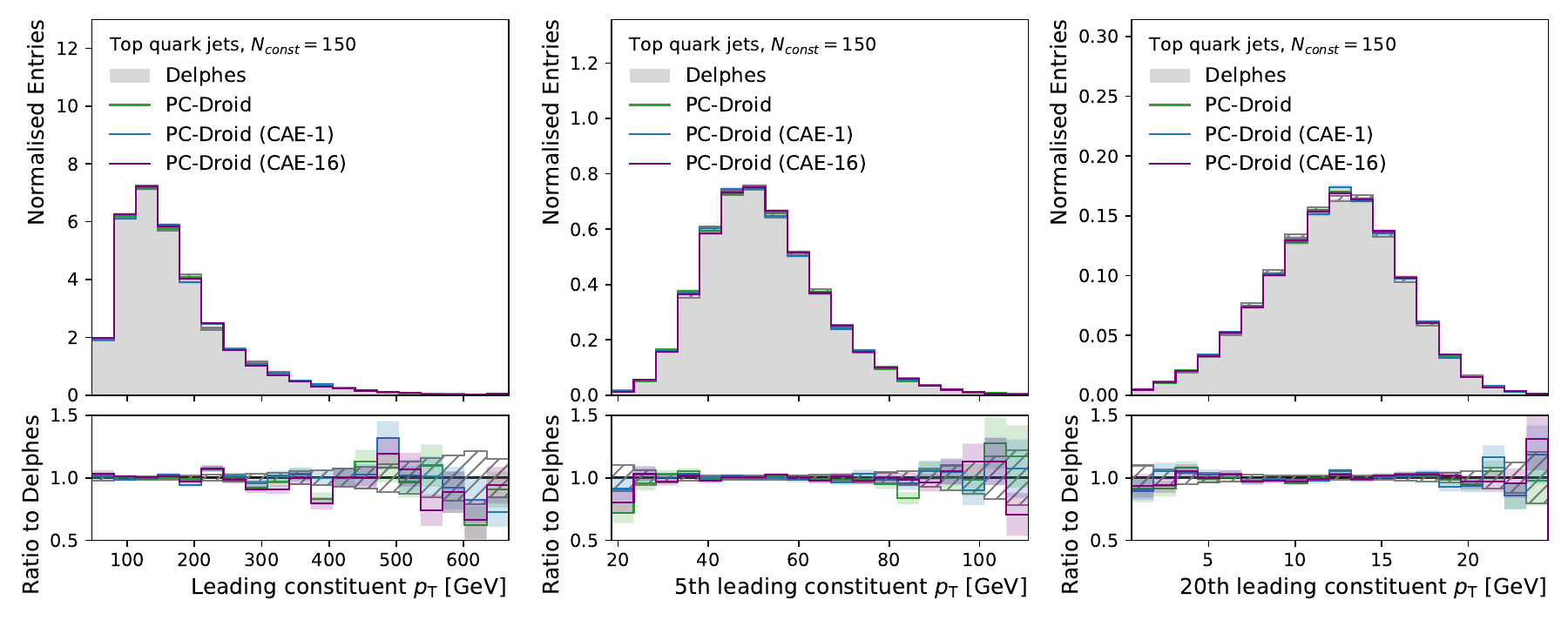} \\
        \includegraphics[width=0.9\linewidth]{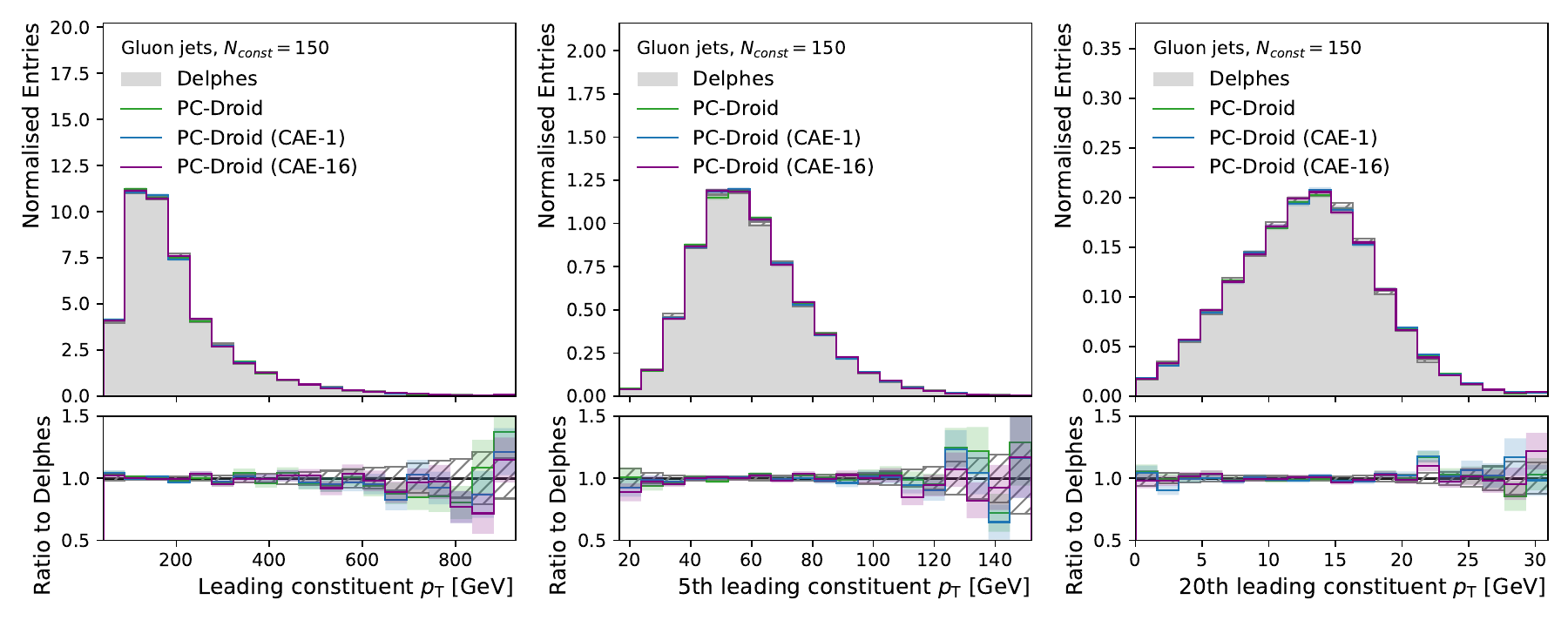}
    \caption{Comparison of \pt distributions of the leading, fifth leading, and twentieth leading constituents of the generated top and gluon jets with up to 150 constituents.
    }
    \label{fig:const-pt_dist-150}
\end{figure*}


\end{document}